# Beat-frequency-resolved two-dimensional electronic spectroscopy: disentangling vibrational coherences in artificial fluorescent proteins with sub-10-fs visible laser pulses

**MASAAKI TSUBOUCHI,**[1,*] **NOBUHISA ISHII,**[1] **YUJI KAGOTANI,**[2] **RUMI SHIMIZU,**[2] **TAKATOSHI FUJITA,**[2] **MOTOYASU ADACHI,**[2] **AND RYUJI ITAKURA**[1]

[1]*Kansai Photon Science Institute, National Institutes for Quantum Science and Technology (QST), 8-1-7 Umemidai, Kizugawa, Kyoto 619-0215, Japan*

[2]*Institute for Quantum Life Science, National Institutes for Quantum Science and Technology (QST), 4-9-1 Anagawa, Inage, Chiba 263-8555, Japan*

*\*tsubouchi.masaaki@qst.go.jp*

**Abstract:** We perform a beat-frequency-resolved analysis for two-dimensional electronic spectroscopy using a high-speed and stable 2D electronic spectrometer and few-cycle visible laser pulses to disentangle the vibrational coherences in an artificial fluorescent protein. We develop a highly stable ultrashort light source that generates 5.3-fs visible pulses with a pulse energy of 4.7 µJ at a repetition rate of 10 kHz using multi-plate pulse compression and laser filamentation in a gas cell. The above-5.3-fs laser pulses together with a high-speed multichannel detector enable us to measure a series of 2D electronic spectra, which are resolved in terms of beat frequency related to vibrational coherence. We successfully extract the discrete vibrational peaks behind the inhomogeneous broadening in the absorption spectra and the vibrational quantum beats of the excited electronic state behind the strong stationary signal in the typical 2D electronic spectra.

## 1. Introduction

When multiple quantum states are simultaneously excited by broadband coherent light, coherent superposition states are generated. These superposition states evolve as a wave packet until coherence is lost through decoherence processes, *e.g.*, collision, spontaneous emission, and so on. This phenomenon is referred to as "quantum coherence". Usually, coherence is identified as a quantum beat signal whose frequency is the inverse of the energy splitting between the coherent superposition states. Following the development of ultrashort pulse lasers and advances in the temporal resolution of spectroscopic tools, vibrational coherence has been detected for molecules in solution prior to phase relaxation due to collisions with solvent molecules, and then analyzed under the formalism of nonlinear optical spectroscopy [1]. The existence of quantum coherence in photosynthetic proteins in solutions has also been recently reported and actively discussed [2, 3].

A pair of photoexcited adjacent chromophores can form exciton states. When the chromophores are excited by an ultrashort laser pulse whose spectrum is broad enough to cover the split energy levels of the exciton states, quantum coherence is expected to be generated in the pair of exciton states. It has been suggested that quantum coherence in photosynthetic proteins is related to the unidirectional energy flow with extremely high efficiency occurring in the photosynthesis process [4]. The long-lived quantum coherence of a pair of exciton states has been reported for the Fenna–Matthews–Olson complex [5-7], phycocyanin [8-11], and allophycocyanin [12, 13] with a phase-relaxation time longer than several hundred femtoseconds. On the other hand, there are strong claims that the detected coherence is not due to excitonic coherence but rather to the vibronic coherence within a single chromophore [3, 14]. The occurrence of quantum coherence during photosynthesis has not yet been confirmed.

Two-dimensional electronic spectroscopy (2D-ES) was employed to evaluate this problem in the studies mentioned above. It can spectrally resolve quantum states that are not able to be resolved using conventional spectroscopies, owing to the inhomogeneous broadening that occurs in solution, and it can also map the correlation between the states [3, 7, 9, 15-21]. Measurement of the correlation map as a function of the pump-probe delay time reveals a reaction pathway from a specific state excited by an initial pump pulse. However, in the time-dependent 2D electronic spectra, the quantum beat signal between the superposition states can hardly be seen because it is usually hidden by the strong stationary signal, which is not related to the temporal evolution of the coherently excited superposition states but rather with that of the single quantum state. To overcome this difficulty in the 2D-ES, beat-frequency-resolved analysis of 2D electronic spectra has been proposed to illustrate the vibronic and exciton states and the coherences between them as coherence amplitude maps [22-24].

In this study, we demonstrate the visualization of the vibrational coherences in an artificial protein with beat-frequency-resolved 2D-ES. We successfully extract the vibrational coherences behind the strong stationary signal that dominates typical 2D spectra and the inhomogeneous broadening in the absorption spectra. First, we introduce a stable and intense sub-10-fs visible light source using a 10-kHz Yb:KGW laser combined with two-stage pulse compression. Second, a high-acquisition-speed 2D electronic spectrometer is built by employing a line-scan camera synchronized to the high-repetition rate of the laser pulse. Finally, we employ beat-frequency-resolved 2D-ES to measure the vibrational coherences in the artificial fluorescent protein RasM, which is synthesized through genetic recombination.

## 2. Experimental Procedure

### 2.1. Sub-10-fs visible pulse generation

We produce sub-10-fs visible pulses using a Yb:KGW laser (LIGHT CONVERSION, PHAROS-SP-10W-200kHz, 1030 nm, 1.0 mJ, 10 kHz, 190 fs) followed by two pulse compression stages, as shown in Supplemental Fig. S1(a). The first pulse compression stage employs a multi-plate pulse compression (MPC) technique [25]. A partial output (0.8 mJ) from the Yb:KGW laser is loosely focused into a gas cell filled with 1.0 bar of helium to avoid nonlinear effects around the focus. The exit fused-silica window of the gas cell and the following second and third fused-silica plates act as thin plates to achieve spectral broadening in our MPC setup. The output pulses are temporally compressed by two Gires–Tournois interferometer mirrors down to 40–50 fs. The spectra before and after the MPC stage are shown in Supplemental Fig. S1(b).

The second pulse compression stage employs spectral broadening based on laser filamentation [26]. The output pulses from the first MPC stage are focused by a concave mirror into a mixed gas cell (0.5 bar of xenon and 0.5 bar of helium), leading to the generation of white light via laser-based filamentation. The visible spectral part of the white light is selected out by a shortpass filter. The transmitted visible pulses are recollimated and compressed by two chirp-mirror pairs. The white light spectrum after the pulse compression is depicted in Fig. 1(a).

After fine-tuning the residual dispersion using a pair of fused-silica wedge plates, we characterize the compressed visible pulses using a second-harmonic-generation frequency-resolved optical gating (FROG) apparatus. The FROG results are summarized in Supplemental Fig. S2. The full width at half maximum of the measured pulse duration is 5.3 fs with a FROG error of 0.85%, as shown in Fig. 1(b). The pulse energy after the pulse compression section is approximately 4.7 µJ. The long-term power fluctuation is measured to be 0.26% (standard deviation) for 25 hours.

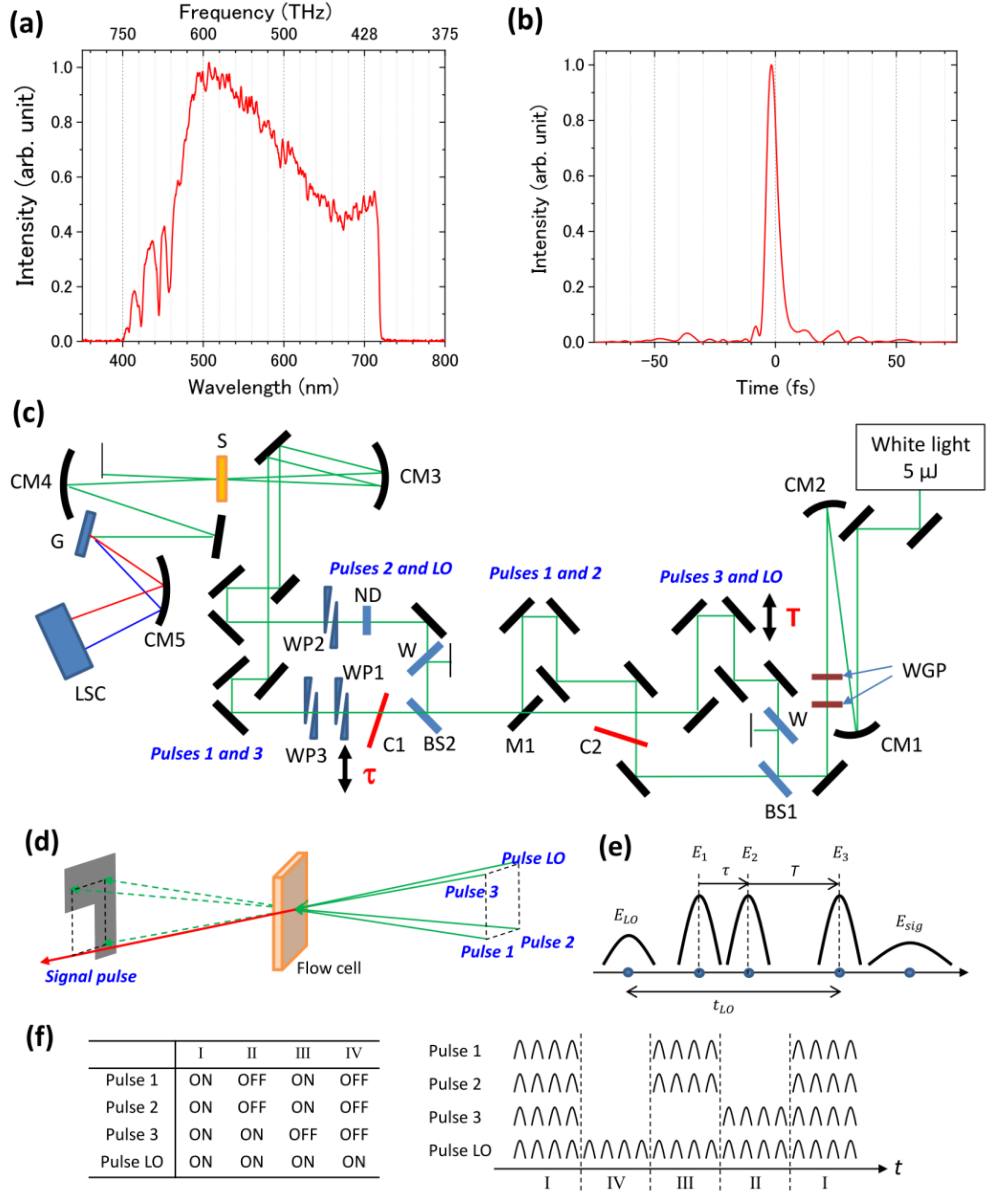

Fig. 1. (a) Spectrum and (b) temporal profile of the sub-10-fs visible pulse output after chirp-mirror-pair-based pulse compression. (c) Schematic diagram of the 2D-ES apparatus. CM 1, 2, 3, 4, and 5: concave mirrors with focal lengths of 150, 75, 250, 250, and 250 mm, respectively; WGP: wire grid polarizer for adjusting the pulse energy; BS: 1-mm-thick fused-silica 1:1 beam splitter; W: 1-mm-thick fused-silica window for dispersion compensation; C1 and 2: optical choppers; M1: silver-coated mirror reflecting only pulses 1 and 2; ND: neutral density filter; WP 1, 2, and 3: fused-silica dispersion wedge plates with a wedge angle of 2.8°; S: sample; G: holographic grating with a groove density of 300 grooves/mm; LSC: high-speed line-scan camera with 4096 pixels. Other thick black lines represent the silver-coated mirrors. The three-dimensional optical layout is shown in Supplemental Fig. S3. (d) BOXCARS geometry of the four-wave-mixing process. (e) Time ordering of the four input pulses and signal light. $E_1$ and $E_2$ are the pump double pulses with a time interval of $\tau$. $E_3$ is the probe pulse delayed from $E_2$ by time $T$. The local oscillator pulse $E_{LO}$ is prior to the pump and probe pulses with the fixed time interval, $t_{LO}$, between $E_{LO}$ and $E_3$. (f) Operation of the double chopping system. The left table summarizes the four cases where pulses 1, 2, 3, and LO are chopped (OFF) or not (ON). The right figure describes the pulse sequences irradiating the sample.

### 2.2. High-speed 2D-ES measurement system

The 2D-ES measurement system is an extended method of 2D infrared spectroscopy [27-29] to the visible spectral region to investigate the electronic state of molecules. The basic ideas of 2D-ES are summarized in References [7, 17]. The experimental setup employed in this study is shown in Fig. 1(c). The sub-10-fs laser pulse with a wavelength range of 470–710 nm is collimated by a pair of concave mirrors to a beam diameter of 3 mm. The laser pulse passes through a pair of 0.7-mm-thick wire-grid polarizers to optimize the pulse energy for the 2D-ES measurement. The laser pulses in the apparatus are vertically polarized.

In the 2D-ES measurement process, three identical pulses (1, 2, and 3) and one weak local oscillator (LO) are focused on a sample in the BOXCARS geometry shown in Fig. 1(d). The signal light from the four-wave-mixing (FWM) process is spatially separated from pulses 1, 2, and 3, and interfered with the LO pulse to achieve heterodyne detection. First, the laser pulse is split into two identical pulses by the first beam splitter (BS1). One of the pulses passes through the computer-controlled translational stage to scan the temporal delay, $T$, between pulses 2 and 3. The other pulse is made 10 mm lower in height and reflected by the mirror M1 to propagate parallel to the former beam. The second beam splitter (BS2) is used to prepare four identical pulses. Two glass plates (W) are used to give identical dispersion to the transmission of the beam splitters. The temporal delay, $\tau$, between pulses 1 and 2 is scanned (or varied) by a pair of dispersion wedge plates (WP1) with a wedge angle of 2.8° in the optical path of pulse 1, as described below. The other two pairs of wedge plates (WP2 and WP3) are inserted into the optical paths of pulses 2 and 3, respectively, to realize identical dispersion for all three pulses. The LO pulse attenuated by the ND filter irradiates the sample at approximately 500 fs prior to the probe (pulse 3). The three-dimensional optical layout around the wedge plates is illustrated in Supplemental Fig. S3.

The four pulses are focused on the sample by a concave mirror (CM3) with a focal length of 250 mm in the BOXCARS geometry shown in Fig. 1(d). Pulses 1, 2, and 3 are spatially filtered, and the FWM signal and the LO light are collimated by a concave mirror (CM4) and introduced to a homemade grating spectrometer. The signal and LO pulses are spatially dispersed via a holographic grating with a groove density of 300 grooves/mm and focused on a line-scan CMOS camera (UNiiQA+, e2v) with 4096 pixels.

The time ordering of the pulses and the time interval between the pulses is presented in Fig. 1(e). In the 2D-ES measurement process, the molecules are excited by the double pulses 1 and 2 with the time interval, $\tau$, and probed by pulse 3 delayed by time $T$ after pulse 2. The beating signal due to coherent excitation of a pair of vibronic states is measured by scanning the waiting time, $T$. The double pump pulses are used to obtain the excitation spectra. We measure the time-domain interferogram by scanning the time interval, $\tau$, and obtain the spectrum from the Fourier transformation of the interferogram. Because visible light with a center wavelength of 550 nm (a period of 1.8 fs) is used in this study, the sub-femtosecond temporal resolution of the scan of $\tau$ is required for the measurement of the interferogram. High temporal resolution is achieved using the pair of dispersion wedge plates [30, 31] as described in Supplemental Fig. S4.

The line-scan camera is synchronized to the laser pulse with a repetition rate of 10 kHz. The line images are transferred to a laptop computer via the 5 GBASE-T system to achieve high-speed acquisition of the spectra of the heterodyne signal. To achieve a background-free heterodyne signal, a dual chopping system is employed with two choppers C1 and C2, as shown in Fig. 1(c) [32-34]. Choppers C1 and C2 are synchronized to the laser pulse and operated at frequencies of 625 Hz and 1250 Hz, respectively. C1 chops pulse 3 and C2 chops pulses 1 and 2. The spectra are averaged for four laser shots, and recorded to Bins I, II, III, or IV according to the chopper timing, as shown in Fig. 1(f). To subtract the spectra of the scattered light of pulses 1, 2, and 3, and the transmitted LO light, the spectrum of S = I – II – III + IV is calculated. This operation is repeated thirty times at each delay time of $\tau$ and $T$, and the thirty spectra of S are averaged. The time interval, $\tau$, is scanned from −60 to 60 fs in steps of 0.4 fs, and the waiting time, $T$, is scanned from 0 to 1000 fs in steps of 4 fs. In total, 480 spectra are measured for each delay time $\tau$ and $T$, and, therefore, a complete set of $3.6 \times 10^7$ (= $301 \times 251 \times 480$) spectra are obtained from the entire course of measurement. Including the waiting time for stage movement, the measurement is completed in seven hours.

In our apparatus, the beam splitters are used to prepare identical pulses instead of diffractive optics (DO), as employed in a previous study [7]. Although DO can reduce the source of phase instability during scanning with sub-femtosecond precision, the broadband spectra used in this study are too dispersed in space and time by DO to be compressed. The dispersion of the laser pulse at the sample position is compensated for by changing the number of bounces on the chirp mirrors and the insertion of the additional wage plates placed prior to the 2D-ES apparatus. The shortest pulse width is evaluated to be < 8 fs by the transient-gating FROG using a 1-mm-thick fused-silica plate at the sample position. The standard deviation of the phase fluctuation is estimated to be $\lambda/76$ for the short term within 10 minutes and $\lambda/23$ for the long term over 10 hours as described in Supplemental Section 3.

### 3. Application of 2D-ES: vibrational coherence in RasM

#### 3.1. Time-resolved two-dimensional electronic spectra

We apply our 2D-ES apparatus to the measurement of the vibrational coherence in the RasM monomer. The detail of the sample is described in Supplemental Section 4. The input pulse energy is 30 nJ/pulse, which does not induce multiphoton excitation to higher excited states by a single pulse.

Figure 2(a) shows the 2D power spectrum of the rephasing signal measured at the waiting time $T$ of 40 fs. In the 2D spectra shown in this paper, we do not plot the amplitude of the real and imaginary parts derived using the phasing procedure but plot instead the intensity, which is the square of the electric field of the signal light. The FWM signal mainly appears in the upper left region from the diagonal line, which means that the detection energy $E_{det}$ is higher than the excitation energy $E_{exc}$. The dashed lines indicate the frequencies of the peak ($E_{exc}$ = 512 THz) and the shoulder ($E_{exc}$ = 550 THz) on the absorption spectrum of RasM shown in Supplemental Fig. S7(c). Fig. 2(b) and (c) show the projection of the 2D spectrum onto the excitation and detection frequency axes, respectively. In the projection of the 2D spectrum on the excitation frequency axis, the signal is observed at the peak and shoulder frequencies. The projected spectrum shows good agreement with the absorption spectrum shown in the same figure except for the long tail at frequencies lower than 500 THz. On the other hand, the broadband spectrum is found from $E_{det}$ = 510 to 640 THz on the detection frequency axis which has a frequency shift of +50 THz as compared to the absorption spectrum.

In most previous 2D-ES experiments, the diagonal component is dominant as compared to the off-diagonal component, unlike the observed 2D spectra shown in Fig. 2(a). Figure 2(d) and (e) show the FWM pathways for the rephasing signal in two- and three-electronic-state models, respectively. First, the molecule interacts with a double-pulse pair $E_1$ and $E_2$, and evolves on the electronic ground $|g\rangle$ or excited state $|e\rangle$. After waiting time $T$, the third pulse $E_3$ probes the molecule, and the signal light $E_{sig}$ of the FWM process is detected. The frequency of pulses $E_1$ and $E_{sig}$ correspond to that on the excitation ($E_{exc}$) and detection ($E_{det}$) axes, respectively. When we consider the two-electronic-state model (Fig. 2(d)), the ground state breaching (GSB) and stimulated emission (SE) processes could have contributed to the diagonal peaks in the 2D spectra [1, 35]. On the other hand, the off-diagonal peak at $E_{exc} < E_{det}$ cannot be explained by the two-electronic-state model, and the three-electronic-state model that includes GSB and the excited state absorption (ESA) processes needs to be considered in the analysis (Fig. 2(e)). The GSB process requires strong absorption at both $E_{exc}$ and $E_{det}$, but the absorption spectrum of RasM does not show high optical density at > 550 THz as compared to that at 512 THz. Therefore, it is unlikely that the off-diagonal GSB signal in the three-level system is much stronger than the diagonal one in the two-level system. Therefore, it is probable that the strong component in the observed 2D spectra for RasM is contributed by the ESA scheme. To confirm this, the transition absorption (TA) spectrum is measured and is shown in Supplementary Fig.

S9. A significant ESA signal appears in the TA spectra at 550–650 THz, which is close to the broad peak in the spectra projected on the detection axis shown in Fig. 2(c). The TA spectrum indicates that the transition intensity in the ESA $|f\rangle \leftarrow |e\rangle$ is much stronger than that in the ground state absorption $|e\rangle \leftarrow |g\rangle$ in this probe frequency region (550–650 THz).

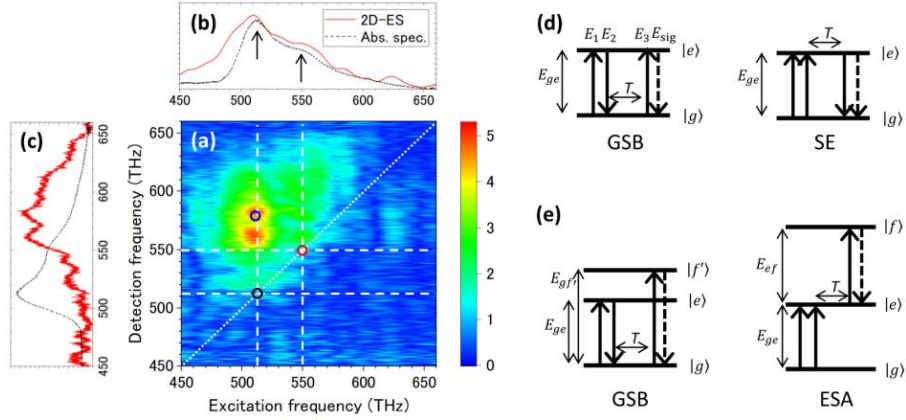

Fig. 2. (a) 2D power spectrum of the rephasing signal measured at the waiting time $T$ of 40 fs. The white dashed lines indicate the frequencies of the peak (512 THz) and shoulder (550 THz) on the absorption spectrum of RasM shown in Supplemental Fig. S7(c). The black, red, and blue circles are plotted at the pair of excitation and detection frequencies: $(E_{\text{exc}}, E_{\text{det}})$ = (512 THz, 512 THz), (550 THz, 550 THz), and (510 THz, 580 THz), respectively. The pump-probe time profiles at these three frequencies are shown in Fig. 3(a). In the (b) top and (c) left panels, the spectra respectively projected onto the excitation and detection frequency axes are shown as red lines. The absorption spectrum of RasM is also shown in (b) and (c) as the black dashed lines. The arrows in (b) indicate the peak and shoulder frequencies of the absorption spectrum. In the main text, the 2D spectrum is explained with the rephasing pathways of the four-wave mixing process in the (d) two-electronic-states model and (e) three-electronic-states model.

### 3.2. Beat-frequency-resolved 2D-ES

Figure 3(a) shows the time profiles of the signal intensities at the diagonal points of $(E_{\text{exc}}, E_{\text{det}})$ = (512 THz, 512 THz) and (550 THz, 550 THz) and at the off-diagonal point of $(E_{\text{exc}}, E_{\text{det}})$ = (510 THz, 580 THz) in the 2D electronic spectra as a function of waiting time $T$. At the diagonal points, the fast oscillations that survives until $T$ = 600 fs are clearly seen. On the other hand, at the off-diagonal point, significant fast oscillation is not observed. Figure 3(b) shows the Fourier-transformed spectra of the time-profiles. The beat frequencies can be identified as $E_{\text{beat}}$ = 35.2 THz for the time-profile measured at $(E_{\text{exc}}, E_{\text{det}})$ = (512 THz, 512 THz), $E_{\text{beat}}$ = 46.0 THz at (550 THz, 550 THz), and $E_{\text{beat}}$ = 16.5 THz at (510 THz, 580 THz). These beating signals are clear evidence of the vibrational coherences. As explained in Fig. 3(c), vibrational coherence is realized at the diagonal peak in the electronic ground and excited states in the GSB and the SE processes, respectively. The beat frequency is identical to the energy separation, $E_{01}$ or $E'_{01}$, between the vibronic states on the electronic ground state or the electronically excited state, respectively. On the other hand, the FWM pathway at the off-diagonal peak ($E_{\text{exc}} < E_{\text{det}}$) is the ESA, as discussed in Sec. 3.1, and the beating signal is contributed by the vibrational coherence of the excited electronic state as shown in Fig. 3(d).

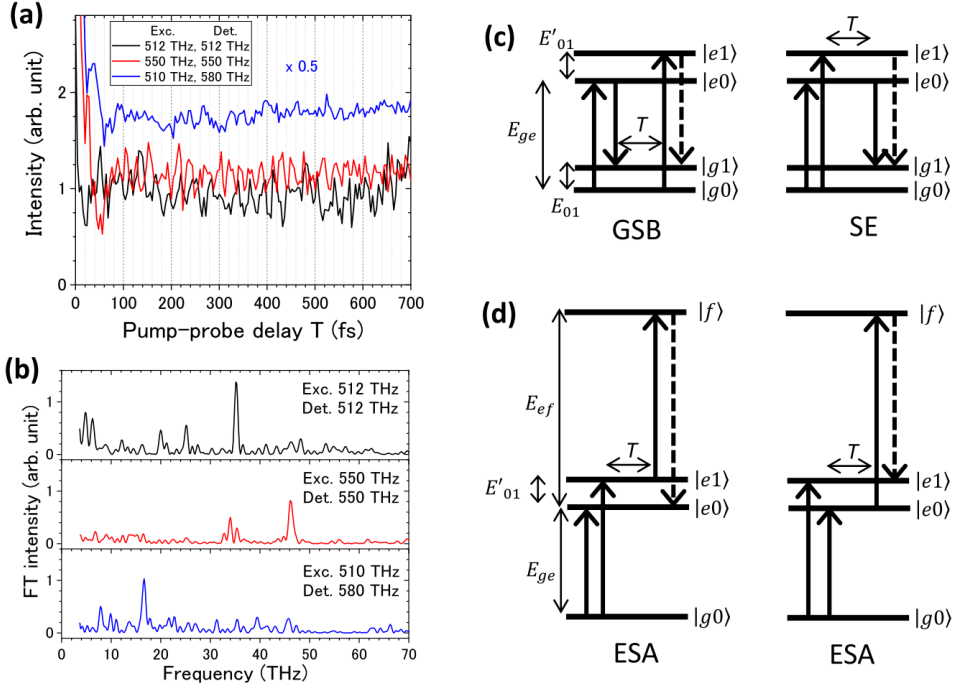

Fig. 3. (a) Pump-probe time profiles observed at the three points of the 2D spectra shown in Fig. 2(a). (b) Beat-frequency spectra calculated by the Fourier transform of the time profile shown in (a). The rephasing pathways of the four-wave mixing process with the vibronic states model for (c) the diagonal peaks and (d) the off-diagonal peak ($E_{exc} < E_{det}$) in the 2D spectra.

To further investigate the vibrational coherence, the FFT spectra of the pump-probe profiles are calculated at all the excitation and detection frequencies on the 2D spectra at intervals of 4 THz. The intensity at the specific beat frequency is mapped as a function of the excitation and detection frequencies. Details of the analysis for the beat-frequency resolved 2D-ES are described in Supplemental Section 9. Figure 4 shows the 2D intensity maps at $E_{beat}$ = 16.5, 25.0, 34.0, 35.2, and 46.0 THz found in Fig. 3(b). In the beat-frequency-resolved 2D spectra, we find a lot of the peaks hidden by the strong stationary signal ($E_{beat} \approx 0$ THz) at approximately ($E_{exc}$, $E_{det}$) = (510 THz, 580 THz) in the typical 2D electronic spectrum shown in Fig. 2(a). Supplemental Table 2 lists the position and width of the peaks found in the beat-frequency-resolved 2D maps. We emphasize that a significant difference is observed between the 2D maps obtained at the beat frequencies of 34.0 and 35.2 THz in Fig. 4(c) and (d), respectively. This difference can also be seen in the beat frequency spectra shown in Supplemental Fig. S10. Two distinguishable peaks (Components 3 and 4 in Supplemental Fig. S10) with an interval of only 1.2 THz (40 cm$^{-1}$) have significantly different dependencies on both the excitation and detection frequencies. Such high beat-frequency resolution is one of the advantages of the present analysis.

In the 2D maps at $E_{beat}$ = 25 and 35.2 THz, the peak A1 is found at the diagonal position of the 2D maps. Both the excitation and detection frequencies at this diagonal position are approximately 512 THz, which is the peak frequency of the absorption spectrum shown in Fig. 2(b). The other diagonal peak, A2, is also found at approximately $E_{exc} = E_{det} \approx 550$ THz, which is close to the shoulder in the absorption spectrum. The weak cross-peaks A3 and A4 are observed at the off-diagonal position ($E_{exc}$, $E_{det}$) = (512 THz, 550 THz) and (550 THz, 512 THz), respectively. The series A peaks can be assigned to the vibrational coherence related to the GSB or SE processes as shown in Supplemental Fig. S11. The Feynman diagrams describing these processes are summarized in Supplemental Fig. S13. The beat frequency $E_{beat}$ is close to the vibrational energies of the electronic ground and excited states, $E_{01}$ and $E_{01}'$, respectively. The vibrational modes with frequencies of 25.0 and 35.2 THz are assigned as $v_2$ and $v_3$, respectively. The vibrational modes are labeled in ascending order of the vibrational energy found in this study and are summarized in Supplemental Table S3.

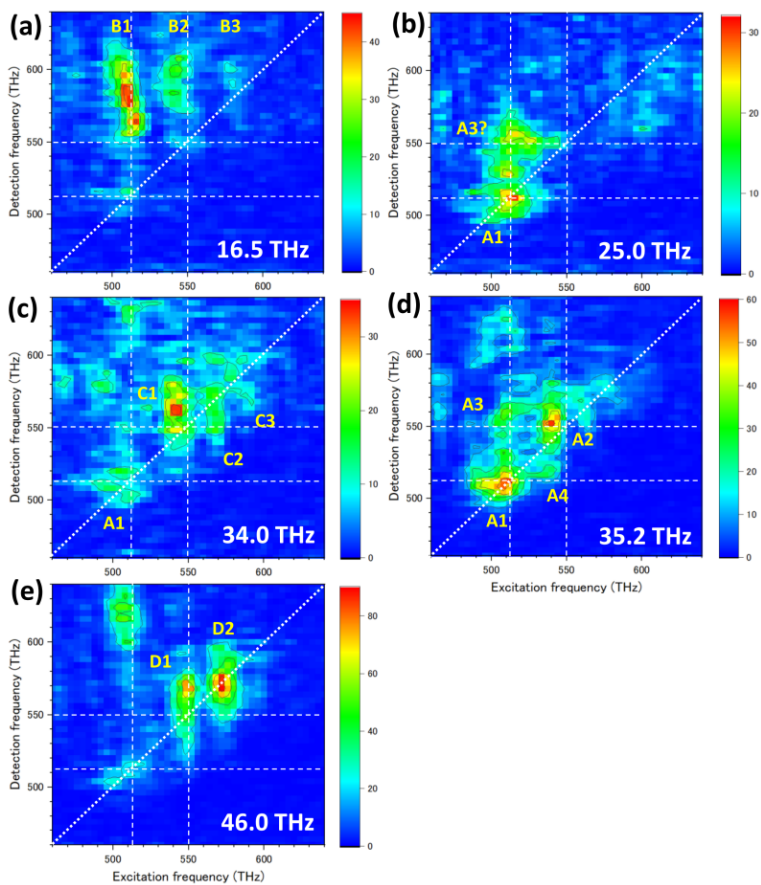

Fig. 4. Beat-frequency-resolved 2D spectra of RasM. The 2D spectra (a), (b), (c), (d), and (e) are extracted from the total 2D spectra at the beat frequencies of $E_{beat}$ = 16.5, 25.0, 34.0, 35.2, and 46.0 ± 1 THz, respectively. The letters A to D correspond to those in Supplemental Table S2.

At $E_{beat}$ = 16.5 THz, the progression of the peaks along the excitation frequency axis is clearly seen in the 2D map. All peaks belonging to the progression are broad along the detection frequency axis. The progression of the peaks along the excitation frequency axis observed in the beat-frequency-resolved 2D-ES is not visible in the absorption spectrum, owing to the occurrence of inhomogeneous broadening. Progression B starts at $E_{exc}$ = 510 THz with an interval of 35 THz on the excitation frequency axis. The frequency of 510 THz is close to the peak frequency in the absorption spectrum, which means that the pump pulse excites the molecules to the bottom of the first electronically excited state $|e\rangle$. The possible pathway for the FWM of progression "B" is shown in Supplemental Fig. S12(a). Here, we assume the ESA process based on the discussion of the TA spectrum as described in Supplemental Section 8. The vibrational coherence with $E_{beat}$ = 16.5 THz is generated for electronically excited state $|e\rangle$. This beat frequency corresponds to the fundamental frequency of the vibrational mode $v_1$. The frequency interval 35 THz of series B along the excitation frequency axis can be assigned to the frequency of the vibrational mode $v_3$. This indicates that the vibrational wave packet associated with the vibrational mode of $v_1$ is induced with the excitation of the vibrational mode $v_3$. At the end of this section, the vibrational modes are tentatively assigned to specific vibrational motion of RasM. After waiting time $T$, the pulse 3 probes the vibrational wave packet via the electronically high-excited state $|f\rangle$, and the signal light is emitted from $|f\rangle$ to $|e\rangle$. The widths in the 2D spectrum of this component are narrow along the excitation frequency axis but broad along the detection axis. This means that the potential energy surface of state $|f\rangle$ may be repulsive or short-lived quasi-bound, whereas the shape of the potential energy surface of the state $|e\rangle$ is similar to that of the state $|g\rangle$.

The same discussion holds for progression "C" on the 2D map at the beat frequency 34.0 THz. The FWM pathways are shown in Supplemental Fig. S12(b). The progression starts at $E_{exc}$ = 542 THz with an interval of 23 THz on the excitation axis. The vibrational wave packet with the beat frequency $E_{beat}$ = 34 THz is generated on $|e\rangle$. The frequency interval 23 THz of series C along the excitation frequency axis can be assigned to the frequency of the vibrational mode $v_2$. This indicates that the vibrational wave packet associated with the vibrational mode of $v_3$ is induced with the excitation of the vibrational mode $v_2$.

In the 2D map at $E_{beat}$ = 46.0 THz, the strong peaks D1 and D2 are seen in the off-diagonal and diagonal positions, respectively. These peaks can be interpreted in terms of two possible ESA pathways, as shown in Supplemental Fig. S12(c). In Case 1, the vibrational mode $v_5$ contributes to the vibrational coherence with $E_{beat}$ = 46 THz for the excited vibronic states $|e, v_2 = 0, v_4 = 1\rangle$ (D1) and $|e, v_2 = 1, v_4 = 1\rangle$ (D2). The vibrational frequency of mode $v_4$ is 37 THz. In Case 2, the second excited electronic state $|e'\rangle$ is considered to be at 37 THz above $|e\rangle$ instead of at vibrational mode $v_4$. To determine the pathway for the series D peaks, further experimental and theoretical investigations are required.

The five observed vibrational modes of the electronically excited states are summarized in Supplemental Table S3. To assign the observed vibrational frequencies to the specific vibrational mode of RasM, vibrational analysis with quantum-chemistry calculations is required. Quantum-chemistry calculation of the electronically excited states of the chromophore in the protein environment is not easy to perform. The chromophore within RasM is a derivative of 4-hydroxy-benzylidene-1,2-dimethylimidazolinone (HBDI), which is the chromophore within the green fluorescent protein (GFP). Isolated HBDI in water was investigated by femtosecond-stimulated Raman spectroscopy, and the vibrational frequencies of the excited electronic state were measured [36]. These frequencies were assigned with the help of quantum-chemistry calculation *in vacuo*. The vibrational frequencies of RasM may not be significantly different from those of HBDI. From this assumption, the vibrational mode $v_2$ may correspond to out-of-plane motions of the hydrogens in the phenolate ring, modes $v_3$ and $v_4$ may correspond to in-plane deformation of the imidazolinone ring, and mode $v_5$ may correspond to in-plane vibration of the phenolate ring. Further theoretical investigation has to be conducted in the vibrational analysis.

## 4. Summary and perspectives

In this study, we performed beat-frequency-resolved two-dimensional electronic spectroscopy (2D-ES) to investigate the vibrational coherences in the artificial fluorescent protein RasM. By employing beat-frequency-resolved analysis, we successfully extracted the vibrational coherence behind the strong stationary signal in the typical 2D spectra and the inhomogeneous broadening in the absorption spectra.

To achieve this, we demonstrated sub-10-fs stable visible pulse generation using a commercial Yb:KGW laser followed by two pulse compression stages: the multi-plate pulse compression technique and spectral broadening based on laser filamentation. A visible pulse with a wavelength region of 470–710 nm, pulse energy of 4.7 µJ, and pulse duration of 5.3 fs was generated at a repetition rate of 10 kHz. The long-term power stability was measured to be 0.26% for 25 hours. By employing the sub-10-fs laser pulse as a light source and a high-speed line-scan camera as the sensor of the spectrometer, the 2D electronic spectrometer could complete the series of measurements within seven hours. The phase stability of the heterodyned four-wave-mixing (FWM) signal was evaluated to be $\lambda/76$ for the short term (10 minutes) and $\lambda/23$ for the long term (over 10 hours).

The four signal components A, B, C, and D associated with the vibrational coherence were extracted from the raw 2D electronic spectra using beat-frequency-resolved analysis and then related to specific FWM pathways. The five vibrational modes were obtained, and tentatively assigned based on the assumption of structural similarity with 4-hydroxy-benzylidene-1,2-dimethylimidazolinone (HBDI). We will extend this method to observe the quantum coherence between the exciton states in an RasM dimer or a hetero dimer of RasM and its derivative. Because we have already assigned the vibrational coherence in the RasM monomer, we can easily distinguish it from the coherence between the exciton states. This method will be applied to natural and artificial photosynthesis proteins to investigate the relationship between the coherence phenomena and the high-efficiency unidirectional energy transfer in photosynthesis.

## Funding

This research was supported by the MEXT Quantum Leap Flagship Program (JPMXS0120330644). We are grateful for the financial support of JSPS KAKENHI (JP21H01898).

## Acknowledgements

The authors thank Y. Yonetani and A. Tanaka at QST for valuable discussions of quantum coherence within biomolecules. We also thank A. Ishizaki at Institute for Molecular Science for valuable comments about 2D-ES.

Supplemental Issues

# Beat-frequency-resolved two-dimensional electronic-spectroscopy: disentangling vibrational coherences in artificial fluorescent proteins with sub-10-fs visible laser pulses

Masaaki Tsubouchi,[a] Nobuhisa Ishii,[a] Yuji Kagotani,[b] Rumi Shimizu,[b] Takatoshi Fujita,[b] Motoyasu Adachi,[b] and Ryuji Itakura[a]

[a]Kansai Photon Science Institute, National Institutes for Quantum Science and Technology (QST), 8-1-7 Umemidai, Kizugawa, Kyoto 619-0215, Japan
[b]Institute for Quantum Life Science, National Institutes for Quantum Science and Technology (QST), 4-9-1 Anagawa, Inage, Chiba 263-8555, Japan

*Author to whom correspondence should be addressed: tsubouchi.masaaki@qst.go.jp

**Contents**

# 1. Laser system for sub-10-fs visible pulse generation

Visible laser pulses with a short duration and a broad bandwidth are required in 2D-ES to excite multiple exciton states coherently. In previous studies of 2D-ES measurements of photosynthetic proteins, Ti:sapphire lasers were used as the primary workhorse with noncollinear optical parametric amplification (NOPA) [1-3] or the laser filamentation technique [4, 5]. However, they have limitations in output power, pulse-to-pulse energy stability, and pointing stability due to their relatively complex laser system that requires a Q-switch laser for pumping. Recently, Yb-based lasers have exhibited much higher output power and more stable output and beam pointing than Ti:sapphire lasers and have also been used to provide 16-fs laser pulses with NOPA for 2D-ES [6].

The experimental setup is summarized in Supplemental Fig. S1. The first pulse compression stage employs a multi-plate pulse compression (MPC) technique [7]. A partial output (0.8 mJ) from the Yb:KGW laser is focused by an AR-coated lens ($f$ = 1000 mm) into a gas cell filled with 1.0 bar of helium to avoid nonlinear effects around the focus. The exit window (1-mm-thick, non-coated fused-silica plate) of the gas cell, which is placed 1120 mm downstream of the position of the focusing lens and normal to the laser propagation direction, acts as the first thin plate in the MPC setup. The second and third 0.5-mm-thick fused-silica plates, which are set at Brewster's angle, are placed 95 mm and 230 mm downstream of the exit window, respectively. After spectral broadening occurred in the first MPC stage, output pulses with an energy of 484 µJ (~60% throughput) are temporally compressed by four bounces on two Gires–Tournois interferometer (GTI) mirrors (Edmund, 12-333, GDD = −500fs$^2$) down to 40–50 fs, dependent on fine adjustment of the MPC configuration. The spectra before and after the MPC stage are shown in Supplemental Fig. S1(b).

The second pulse compression stage employs spectral broadening based on laser filamentation [8]. The output pulses from the first MPC stage are focused by a concave mirror ($f$ = 1500 mm) into a mixed gas cell (0.5 bar of xenon and 0.5 bar of helium), leading to the generation of white light via laser-based filamentation. The visible spectral part of the white light is selected out by a shortpass filter (Edmund, 86-095, transmittance range 400–710 nm). The transmitted visible pulses are recollimated ($f$ = 750 mm) and compressed by two sets of chirp mirror pairs (five bounces on each of a chirp mirror pair (Layertec, 111346) and six bounces on each of the other chirp mirror pair (Layertec, 148545)). The white light spectrum after pulse compression is depicted in Fig. 1(a) in the main text.

After fine-tuning the residual dispersion using a pair of fused-silica wedge plates (Newport, 23RQ12-02-M), we characterize the compressed visible pulses using a second-harmonic-generation frequency-resolved optical gating (FROG) apparatus. The FROG results are summarized in Supplemental Fig. S2. The full width at half maximum of the measured pulse duration is 5.3 fs with a FROG error of 0.85%.

The pulse energy after pulse compression is approximately 4.7 µJ. The long-term power fluctuation is measured to be 0.26% (standard deviation) for 25 hours.

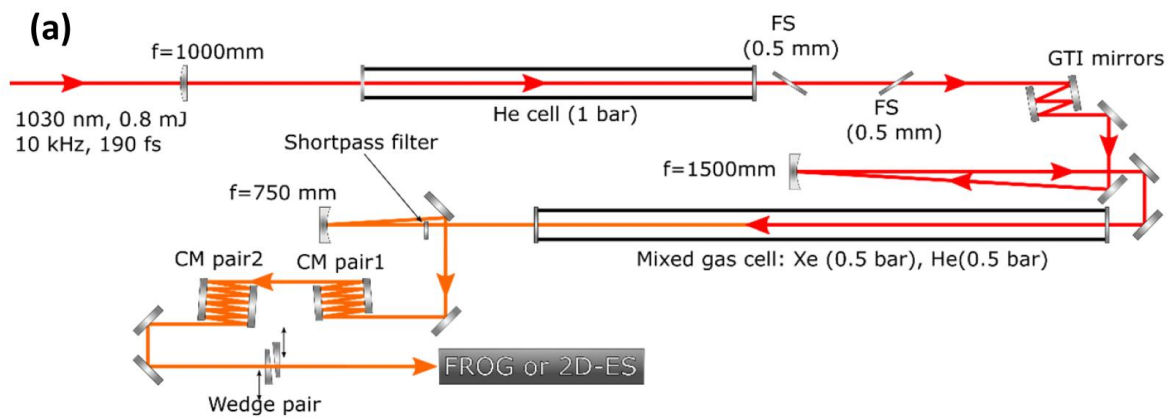

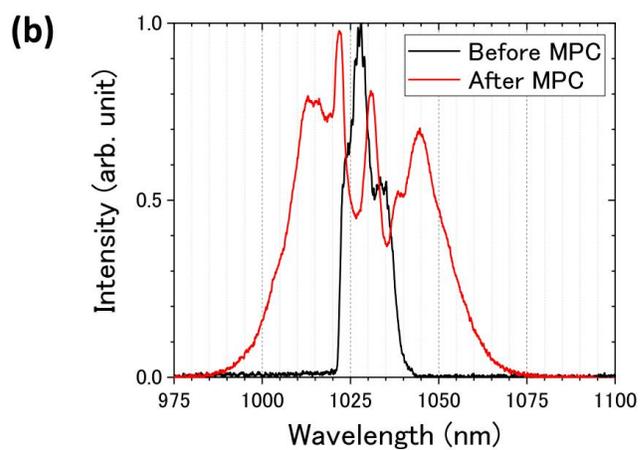

Supplemental Fig. S1. (a) Schematic of the sub-10-fs visible pulse light source. GTI: Gires–Tournois interferometer (GTI) mirror; FS: fused-silica plate. (b) The spectra before (black line) and after (red line) MPC.

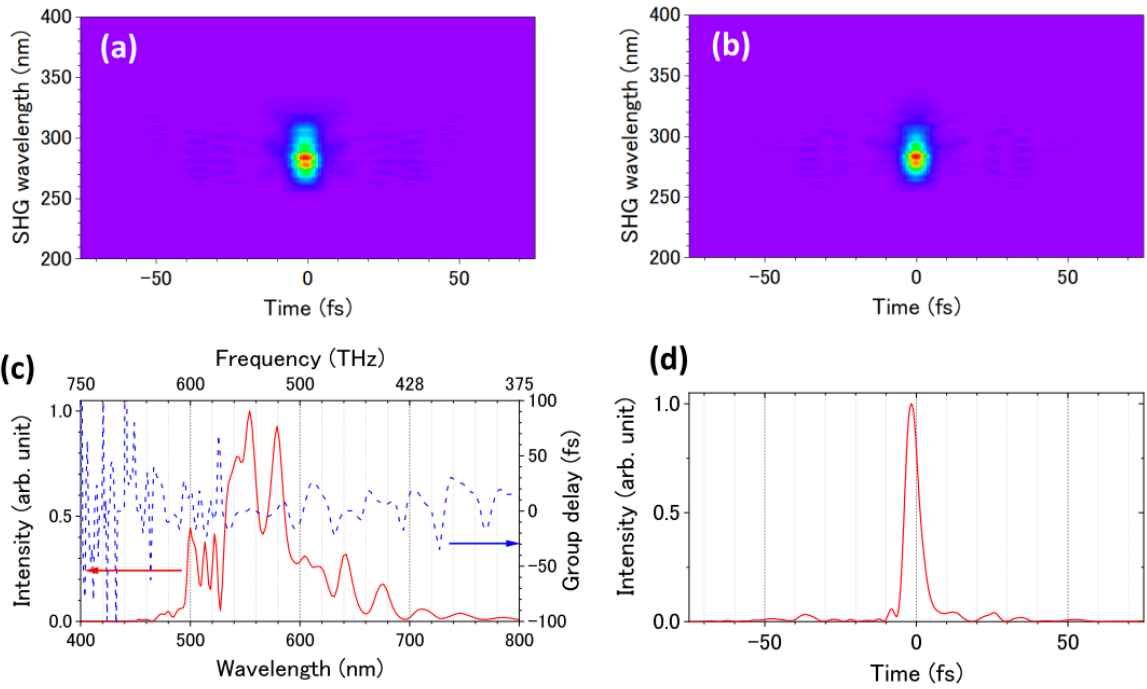

Supplemental Fig. S2. Characterization of the sub-10-fs visible pulses by second-harmonic-generation FROG. (a) Measured and (b) reconstructed FROG traces. (c) Retrieved spectral intensity (red solid line) and residual group delay (blue dashed line). (d) Retrieved temporal profile with a duration (FWHM) of 5.3 fs. The FROG error is 0.85%.

## 2. Temporal scan with sub-femtosecond temporal resolution

In 2D-ES, double pump pulses are used to obtain the excitation spectra. We measure the time-domain interferogram by scanning the time-interval $\tau$ between pulses 1 and 2 (see Fig. 1(e) in the main text) and obtain the spectrum from the Fourier transformation of the interferogram. Because visible light with a center wavelength of 550 nm (a period of 1.8 fs) is used in this study, sub-femtosecond temporal resolution is required in the time scan of $\tau$.

This resolution is achieved using a pair of dispersion wedge plates [9, 10] which is placed as shown in Supplemental Fig. S3. The geometry of the laser pulse propagation and the orientation of the wedge plates is shown in Supplemental Fig. S4(a). When one of the wedge plates is moved by $\Delta x$ perpendicular to the laser pulse propagation, the optical path length of the fused-silica wedge plate increases by $\Delta L$, which is proportional to the increase in the time interval $\Delta \tau$. The relationship between $\Delta x$ and $\Delta \tau$ is calibrated by spectral interferometry of pulses 1 and 2 diffracted from the pinhole placed at the sample position, as described in [10]. Supplemental Fig. S4(b) shows the calibration factor $\Delta x/\Delta \tau$ as a function of the wavelength. Note that the calibration factor relates to the refractive index of the wedge plate. Because broadband laser light is employed in this study, the dispersion of the wedge plate needs to be considered in the calibration and we use the wavelength-dependent calibration factor. The factor is close to 25 μm/fs and, therefore, the temporal resolution of 0.04 fs is achieved via the computer-controlled translational stage with an accuracy of 1 μm.

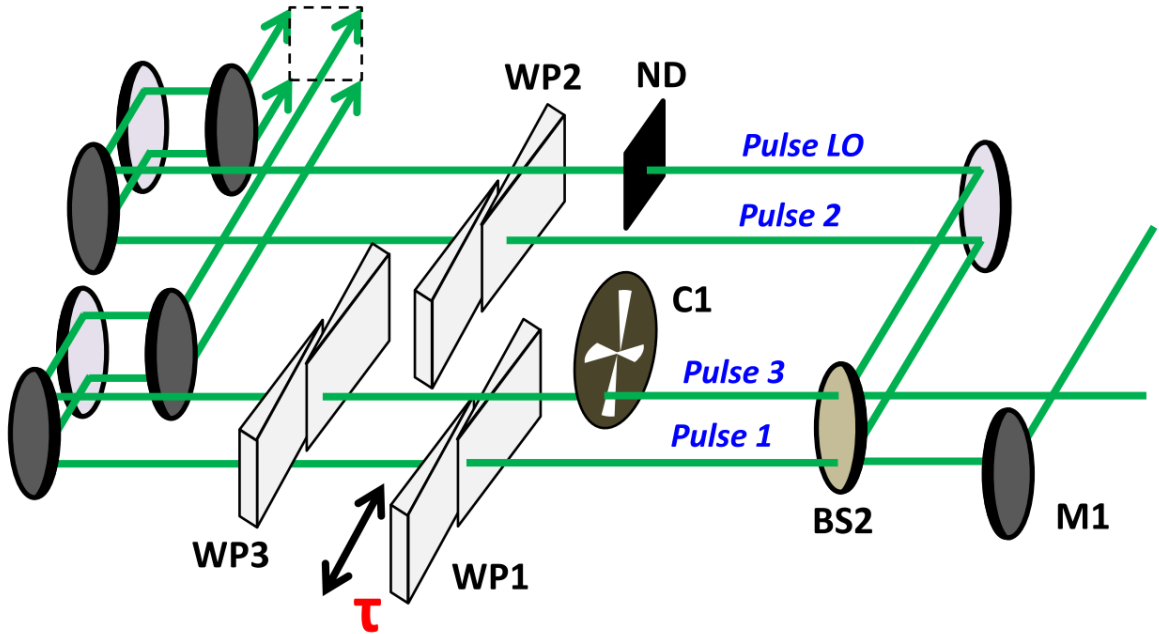

Supplemental Fig. S3. Three-dimensional optical layout for the temporal scan with sub-femtosecond resolution and mechanical chopping. The labels for the optical components correspond to those in Fig. 1(c) in the main text.

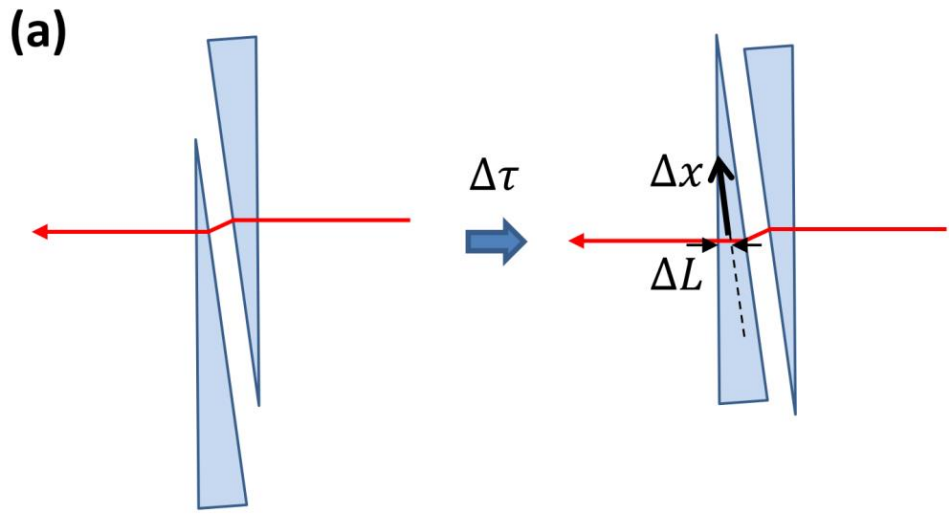

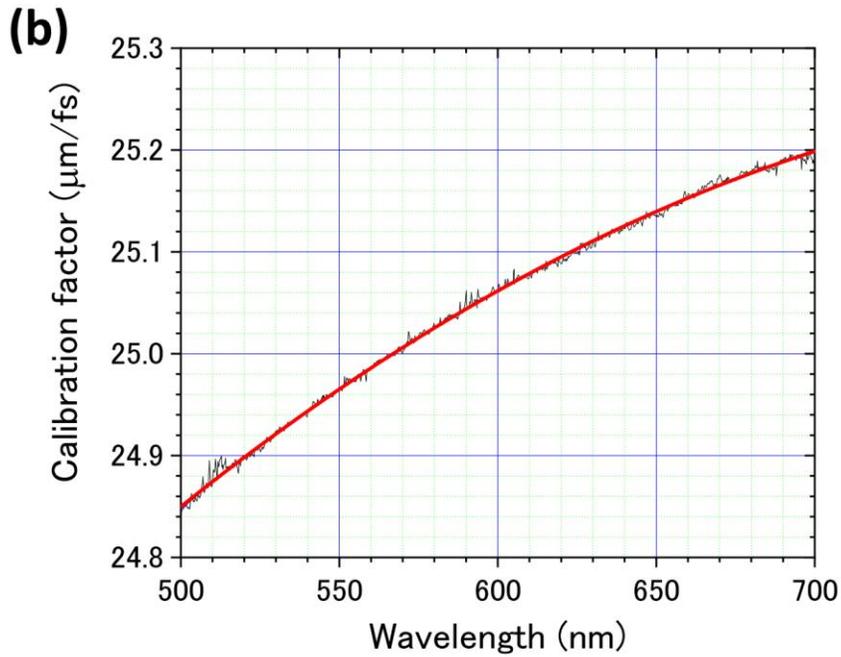

Supplemental Fig. S4. Scanning system for temporal delay with sub-femtosecond resolution and its calibration. (a) Geometry of the laser pulse propagation and orientation of the wedge plates. (b) Calibration factor $\Delta x/\Delta \tau$ as a function of the wavelength.

## 3. Phase stability of the 2D-ES setup

To evaluate the phase stability of our 2D-ES setup, the spectra of the heterodyne signal (S = I – II – III + IV) from the 1-mm-thick fused-silica plate at $\tau = 0$ and $T = 0$ are recorded every 10 seconds over 10 hours and are shown in Supplemental Fig. S5(a). The phase is retrieved from the heterodyne spectrum using the method described in [11]. Supplemental Fig. S5(b) shows the long-term phase stability at the center wavelength of the laser light, 550 nm. The standard deviation of the phase fluctuation is estimated to be $\lambda/76$ for the short term within 10 minutes and $\lambda/23$ for the long term over 10 hours.

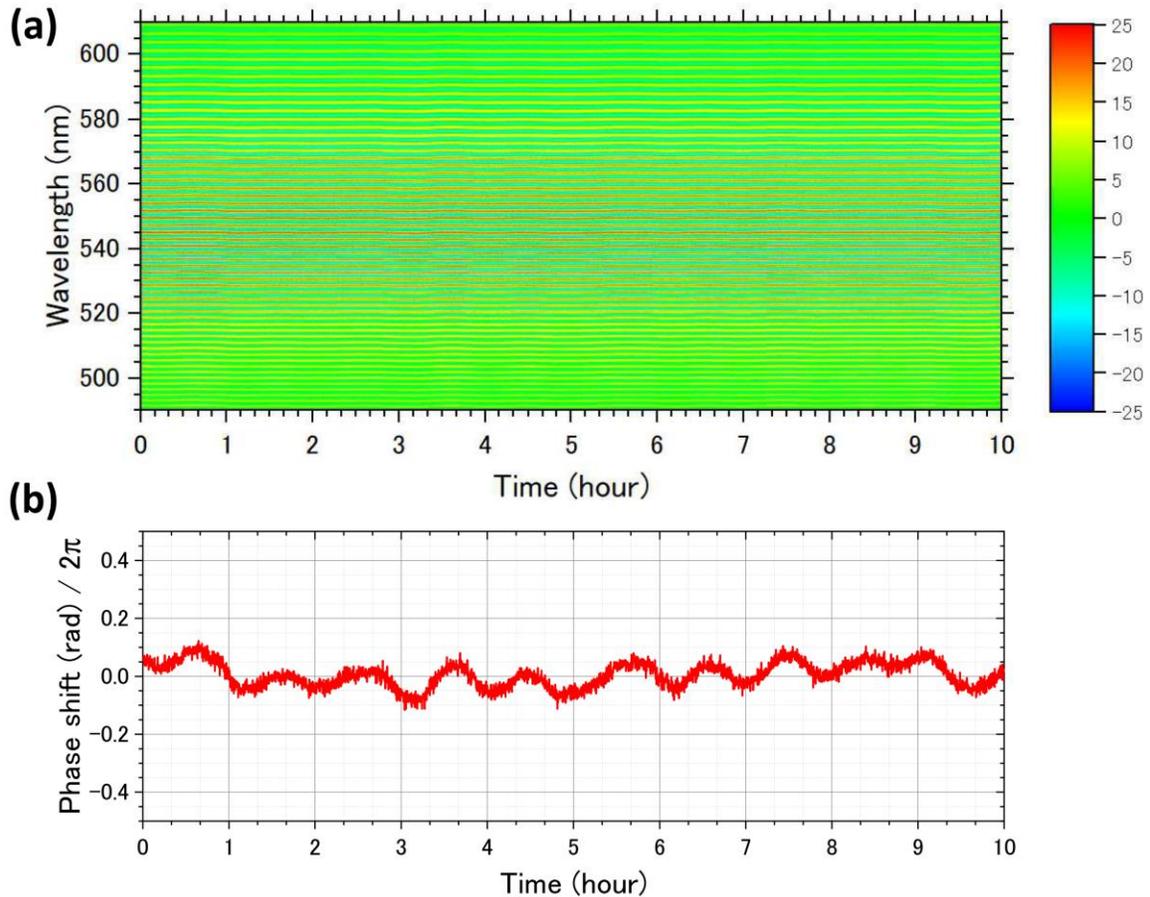

Supplemental Fig. S5. Phase stability of the 2D-ES setup. (a) Long-term measurement of the spectra of the heterodyne signal on the 1-mm-thick fused-silica plate at $\tau = 0$ and $T = 0$ recorded every 10 seconds over 10 hours. (b) Phase at 550 nm retrieved from the heterodyne spectra.

## 4. Artificial fluorescent protein: RasM

It is well known that fluorescent proteins have revolutionized the imaging technologies used in the biological sciences. One of their intriguing properties is that chromophores are post-translationally formed in protein molecules. The variable chromophores are embedded in the barrel structure of the protein and interact with surrounding amino acids with color tuning. A part of the fluorescent proteins form dimers or tetramers in addition to monomers as a ternary structure. Here, we focus on mFruits based on DsRed for the molecular design because of their level of development.

The fluorescent protein RasM is designed and developed based on mRaspberry and mCherry [12-14]. The chemically synthesized DNA that encodes the gene including 255 amino acids is purchased from GENEWIZ (Azenta Life Sciences, Japan), as optimized for *E. coli*. The synthesized DNA is inserted into the pET24a vector using In-Fusion HD Enzyme (Takara Bio, Japan) (Supplemental Fig. S6). The expression plasmid is transfected into *E. coli* BL21(DE3). Supplemental Fig. S7(a) and (b) show the tertiary structure of RasM determined using X-ray crystallography and the chemical structure of the chromophore in RasM, respectively.

The RasM protein is diluted by a buffer solution to prepare the sample solution with a concentration of 5 mg/mL. The protein concentration is adjusted using the TaKaRa BCA Protein Assay Kit (Takara Bio, Japan). The sample solution is circulated by a perista pump through the sample cell with 1-mmthick quartz windows and an optical path length of 0.2 mm. The optical density of the sample solution is OD ~ 0.2 at the peak of the absorption spectrum, as shown in Supplemental Fig. S7(c).

## 5. Sample preparation for RasM

The bacterial cells carrying the plasmid (Supplemental Fig. S6) are grown at 310 K in an LB medium to an OD600 around 0.7, and protein expression is induced at 310 K for 18 hours by adding isopropyl-β-D-thiogalactopyranoside to a final concentration of 0.1 mM. The cells are suspended in a 20 mM Tris-HCl buffer (pH 8.0) containing 0.1 M NaCl, and then disrupted by sonication. The cell lysate is centrifugated at 10,000 g for 30 min. After the centrifugation, the supernatant is dialyzed against the 20 mM Tris-HCl buffer (pH 8.0) at 278 K. After the insoluble materials are removed by centrifugation, the supernatant is applied onto a HiTrap Q XL column with 5 mL volume (Cytiva). The protein is eluted with a linear gradient of 0–1.0 M NaCl in the 20 mM Tris-HCl buffer (pH 8.0). The pooled fraction containing RasM is applied onto a StrepTrap HP column with 5 mL volume (Cytiva) eluted with 2.5 mM desthiobiotin in 100 mM Tris-HCl buffer (pH 8.0), 150 mM NaCl, and 1 mM EDTA. The final yield is 18 mg of RasM protein from 0.2 L of culture.

```
                                          pET24a
                                            ←
                         CGTCCGGCGTAGAGGATCGAGATCTCGATCCCGCGAAACCCCACTCCTCT
          CCCAGTGATTGGAAATATGTGTGTTTAGGAGGAAGTGAGCACATCTTCAGATCTCGATCCCGCGAAATTAATACGACT
          CACTATAGGGGAATTGTGAGCGGATAACAATTCCCCTCTAGAAATAATTTTGTTTAACTTTAAGAAGGAGATATACAT

 -81  ATG AAA TGG AGC CAT CCG CAG TTT GAA AAA AAA ACC GAC AAA ACG GAT AAG ACA GAT AAG   -24
 -27   M   K   W   S   H   P   Q   F   E   K   K   T   D   K   T   D   K   T   D   K    -8

 -23  ACT GAC AAA ACC GAT GAC GAC GAT AAA GGC AGC GCA GCC ATT ATT AAA GAA CAT ATG CGC    39
  -7   T   D   K   T   D   D   D   D   K   G   S   A   A   I   I   K   E   H   M   R    13

  40  TTC AAA GTG CGC ATG GAG GGC AGC GTG AAT GCC CAT GAG TTC GAA ATT GAG GGC GAA GGC    99
  14   F   K   V   R   M   E   G   S   V   N   A   H   E   F   E   I   E   G   E   G    33

 100  GAA GGT CGC CCG TAC GAA GGT ACC CAG ACC GCC AAA CTG AAG GTG ACA AAA GGT GGT CCG   159
  34   E   G   R   P   Y   E   G   T   Q   T   A   K   L   K   V   T   K   G   G   P    53

 160  CTG CCG TTT GCC TGG GAC ATT CTG AGC CCG CAG TTT ATG TAC GGT AGC AAG GCC TAT GTG   219
  54   L   P   F   A   W   D   I   L   S   P   Q   F   M   Y   G   S   K   A   Y   V    73

 220  AAG CAT CCG GCC GAC ATT CCG GAC TAT TTA AAA CTG AGC TTC CCG GAG GGC TTC AAA TGG   279
  74   K   H   P   A   D   I   P   D   Y   L   K   L   S   F   P   E   G   F   K   W    93

 280  GAG CGC GTG ATG AAC TTC GAA GAT GGT GGC GTG GTG ACC GTG ACC CAG GAT AGC AGC CTG   339
  94   E   R   V   M   N   F   E   D   G   G   V   V   T   V   T   Q   D   S   S   L   113

 340  CAG GAC GGC GAG TTT ATC TAC AAG GTG AAG CTG CGC GGT ACC AAT TTC CCG AGC GAT GGC   399
 114   Q   D   G   E   F   I   Y   K   V   K   L   R   G   T   N   F   P   S   D   G   133

 400  CCG GTT ATG CAG AAG AAA ACC ATG GGC TGG GAA GCA AGC AGC GAA CGC ATG TAT CCG GAA   459
 134   P   V   M   Q   K   K   T   M   G   W   E   A   S   S   E   R   M   Y   P   E   153

 460  GAT GGC GCC CTG AAG GGC GAG ATT AAA CAG CGC CTG AAG CTG AAA GAC GGC GGC CAC TAT   519
 154   D   G   A   L   K   G   E   I   K   Q   R   L   K   L   K   D   G   G   H   Y   173

 520  ACT GCC GAG GTG AAA ACC ACC TAC AAG GCA AAG AAA CCG GTG CAG CTG CCG GGC GCA TAC   579
 174   T   A   E   V   K   T   T   Y   K   A   K   K   P   V   Q   L   P   G   A   Y   193

 580  AAA GTG GGT ATC AAA CTG GAC ATC ACC AGC CAC AAC GAG GAC TAC ACC ATC GTG GAA CAG   639
 194   K   V   G   I   K   L   D   I   T   S   H   N   E   D   Y   T   I   V   E   Q   213

 640  TAT GAG CGT GCC GAA GGT CGC CAT AGC CGT CAT CAC CTG TTT GGC TAA                   687
 214   Y   E   R   A   E   G   R   H   S   R   H   H   L   F   G   *                   228

          ACGCCGGATAATCGGCAGCCGAGGAACCGGTAATGAGATCCGGCTAATAACTAGCATAACCCCTTGGGGCCTCTAAACG
          GGTCTTGAGGGGTTTTTTGTGTAGTAATTGTGTAAGTGACTGCATTGGACGCGGATCCGAATTCGAGCTCCGTCG
                                                 ├──→
                                                pET24a
```

Supplemental Fig. S6. Chemically synthesized DNA and amino acid sequences for RasM. The residue number is adjusted to mFruits [15]. The chromophore formed via post-translational modification of the three amino-acid residues is shown in open square. The partial DNA sequences from the vector of pET24a at both sides of RasM are also shown.

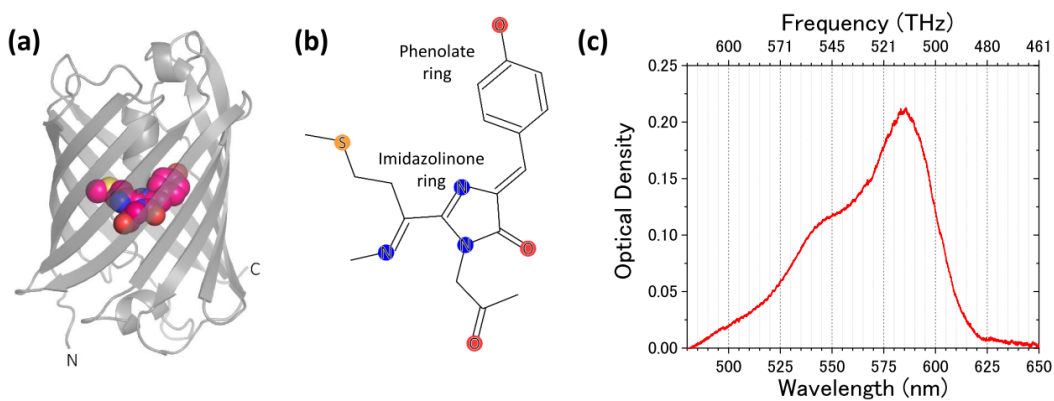

Supplemental Fig. S7. (a) Tertiary structure of RasM determined by X-ray crystallography (see Supplemental Issues). The overall protein molecule is modelled as ribbons. The letters "N" and "C" indicate the N- and C-terminals of the polypeptide chain. A chromophore is shown via space-filling representation. Carbon, oxygen, nitrogen, and sulfur atoms are colored in magenta, red, blue, and yellow, respectively. (b) Chemical structure of the chromophore in RasM. (c) Absorption spectrum of the RasM solution with a concentration of 5 mg/mL and an optical path length of 0.2 mm.

## 6. X-ray crystallography for RasM

For the crystallization of RasM, the DNA coding of 27 amino acids at N-terminal (KWSHPQFEKKTDKTDKTDKTDKTDDDD) is deleted using the PCR method to obtain high-resolution diffraction data. The protein is purified using the HiTrap Q XL column. The crystallization conditions for the RasM are surveyed using the hanging-drop vapor diffusion method with Wizard 3 & 4 HTS (Rigaku Reagents, Inc.) at 298 K. The protein solution of 20 mg/ml (0.4 μL) in 20 mM Tris-HCl buffer (pH 8.0) is mixed with the same volume of reservoir solution. The crystal is obtained using 100 mM HEPES-NaOH buffer (pH 7.5) and 70% (v/v) MPD (No. C4). It is then mounted with a cryoloop and flash-cooled in a nitrogen-gas stream at 100 K at beamline BL5A in the KEK Photon Factory (Tsukuba, Japan). The X-ray diffraction dataset is collected using a monochromatic X-ray beam ($\lambda = 1.00$ Å) at 100 K. The oscillation angle is 0.1° per image and a total of 1800 images are collected. The X-ray diffraction data are integrated and scaled using the XDS program package [16]. Initial phase information for RasM is obtained from the structure previously reported (PDB id: 2H5Q) [15]. The crystallographic refinement is performed using the program PHENIX in X-ray refinement with phenix.refine [17]. The data collection and refinement statistics are summarized in Supplemental Table S1 and the figures are made using the program Pymol [18].

Supplemental Table S1. Data collection and refinement statistics of the neutron and X-ray crystallography.

| Protein | RasM |
|---|---|
| Source | X-ray |
| Solvent | $H_2O$ |
| Collected beamline | BL5A at KEK-PF |
| Wavelength (Å) | 1.000 |
| Temperature (K) | 100 |
| Space group | $P2_12_12_1$ |
| Unit cell parameters | $a = 59.39$ Å<br>$b = 62.33$ Å<br>$c = 107.82$ Å |
| Resolution (Å) | 43.0–1.01 (1.03–1.01) |
| No. of measured reflections | 1,311,250 (59,678) |
| No. of unique reflections | 209,327 (10,267) |
| Redundancy | 6.3 (5.8) |
| $R$-merge (%)† | 5.1 (77.4) |
| Completeness of data (%) | 100 (100) |
| $I$/sig $I$ | 17.9 (2.1) |
| Refinement | |
| $R$-factor (%)‡ | 13.80 |
| $R$-free (%)‡ | 15.81 |
| RMSD bonds (Å) | 0.009 |
| RMSD angles (deg) | 1.315 |
| Mean B value (Å) | 13.76 |
| PDBID | 8GOS |

Values for highest-resolution shells are shown in parentheses.

†$R_{merge} = \Sigma |I - <I>| / \Sigma <I>$, where $I$ is the intensity of the reflection.
‡R-factor and $R_{free} = \Sigma ||F_o| - |F_c|| / \Sigma |F_o|$, where $F_o$ and $F_c$ are the observed and calculated structure-factor amplitudes of the reflection, respectively, and the free reflections (5% of the total used) are held aside for $R_{free}$ throughout refinement.

## 7. Crystal Structure of RasM

The high-resolution crystal structure is determined using a 1.0 Å resolution. The electron density maps are clearly illustrated as shown in Supplemental Fig. S8(a). The chromophore forms hydrogen bonds with three water molecules and three amino acid residues of Arg95, Ser146, and Glu215 (Supplemental Fig. S8(b)). Additionally, it appears that the hydrophobic residues of Pro63, Ile161, Ile197, and Leu199 directly affect conformation of the chromophore. These structures are similar to those of mCherry [15]. However, no significant alternative conformations for Ser46 and Leu199 can be observed that are different to those of mCherry. These results suggest that the structural disorder for RasM is lower than that of mCherry.

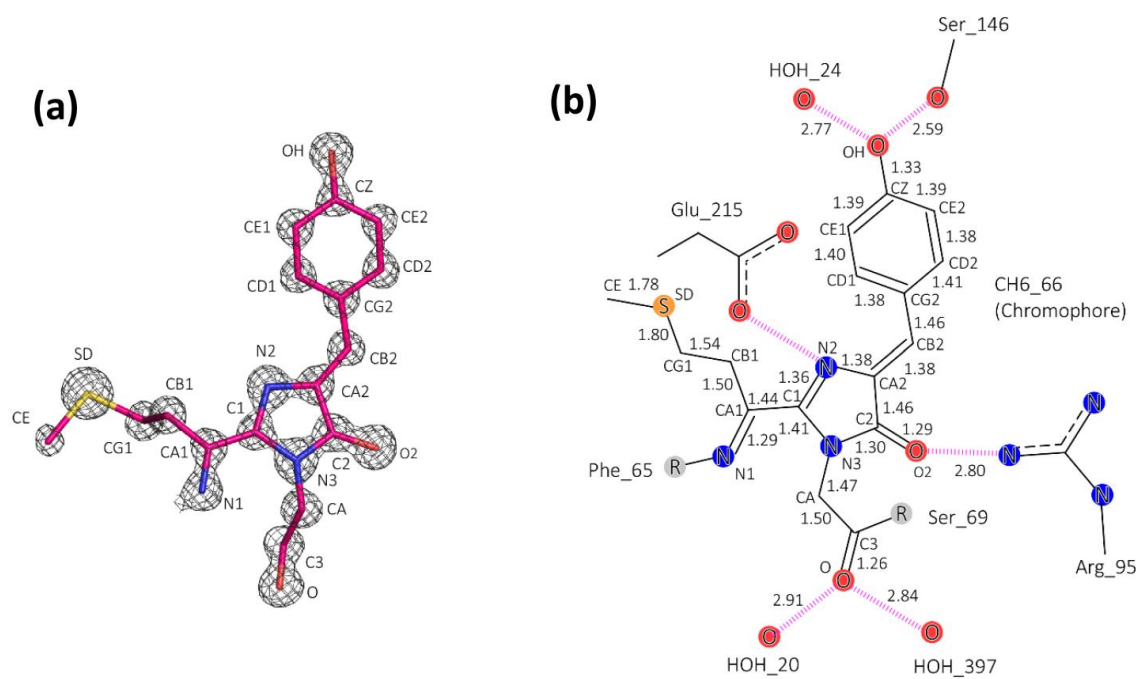

Supplemental Fig. S8. Structural features of the chromophore for RasM. Atom names correspond to those deposited to PDB. (a) Electron density maps of the chromophore drawn at $3\sigma$ level. Carbon, oxygen, nitrogen, and sulfur atoms are shown in magenta, red, blue, and yellow, respectively. (b) Schematic drawing of the chromophore (CH6 66) with three amino acid residues (Arg_95, Ser_146, and Glu_215) and four water molecules (HOH_20, HOH 24, and HOH 397) interacting via hydrogen bonds. Hydrogen bonds are indicated by dotted lines in magenta. The numbers show the bond distances in the chromophore.

## 8. Time-resolved transient absorption spectra

Time-resolved transient absorption (TA) spectra are measured using HELIOS (Ultrafast Systems). The absorption and emission spectra are recorded at room temperature with a model V-760 spectrophotometer (Nihon Bunko) and a model FP-8300 fluorescence spectrometer (Nihon Bunko), respectively.

Supplemental Fig. S9(a) shows the TA spectra as solid lines measured at the excitation wavelengths of 585 nm (frequency of 512 THz) and 545 nm (550 THz), which are the peak and shoulder of the absorption spectrum (blue dash-dot line), respectively. The pump-probe delay time is 500 fs, which is much shorter than the relaxation time of the electronically excited state. In the probe for frequencies lower than 550 THz, the absorbance of the probe light is decreased. Although the peak intensity at 505 THz for the 512 THz excitation is slightly larger than that for the 550 THz excitation, a significant difference is not found between the TA spectra. On the other hand, for probe frequencies higher than 550 THz, the increase of absorption by the 512 THz excitation is twice that by the 550 THz excitation.

The decrease of the absorption in the lower frequency region is due to ground state breaching (GSB) and the stimulated emission (SE) processes. In principle, the shapes of the TA spectrum in the GSB and SE processes are similar to the absorption and fluorescence spectra, respectively. On the other hand, the increase of the absorption in the higher frequency region is due to the excited state absorption process.

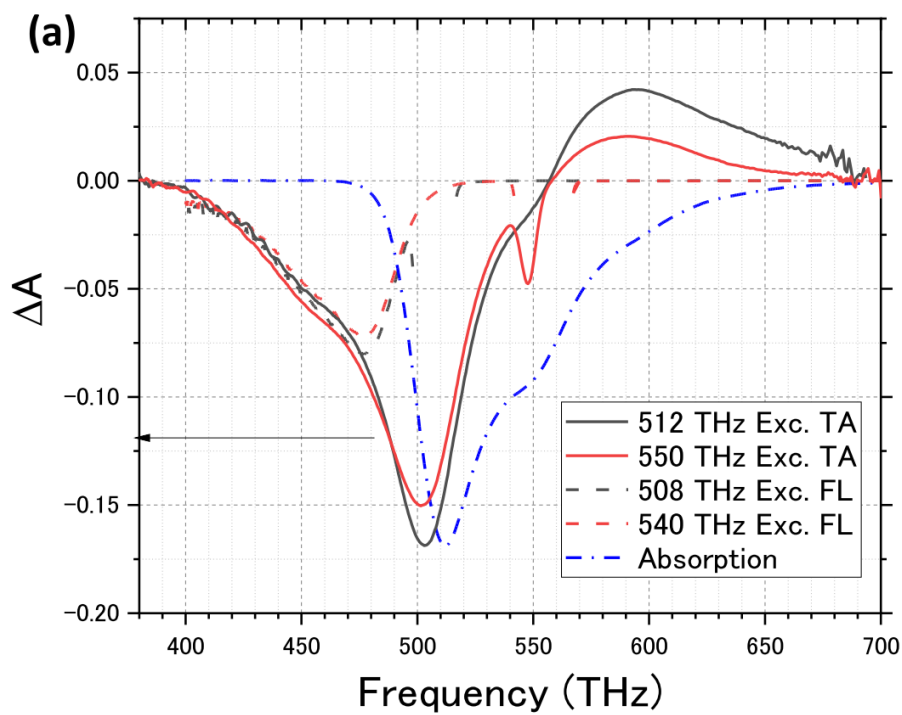

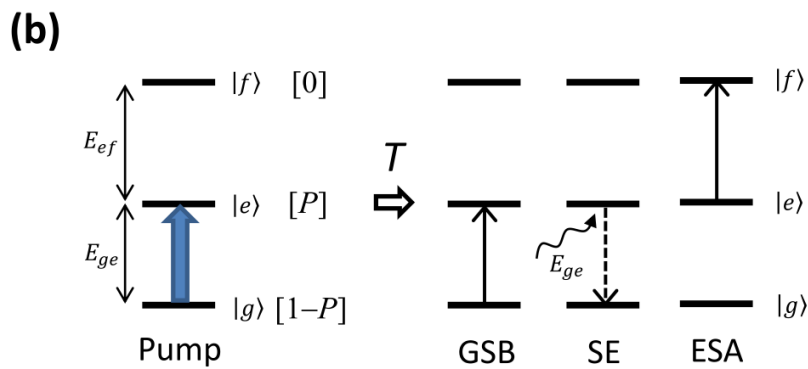

Supplemental Fig. S9. (a) Transient absorption (TA) spectra of RasM compared with the fluorescence (FL) spectra and the absorption spectrum. Black and red solid lines represent the TA spectra measured at the pump wavelength (frequency) of 585 nm (512 THz) and 545 nm (550 THz), respectively. The pump-probe delay time is 500 fs. Black and red dashed lines represent the fluorescence spectra measured at the pump wavelength (frequency) of 590 nm (508 THz) and 540 nm (540 THz), respectively. The blue dash-dot line represents the absorption spectrum. The signs of the fluorescence and absorption spectra have been inverted for comparison. (b) Schematic diagram of TA. GSB, SE, and ESA are considered to be the probe processes. $[1 - P]$, $[P]$, and $[0]$ are the populations of the $|g\rangle$, $|e\rangle$, and $|f\rangle$ just after pump light irradiation, respectively.

## 9. Analysis for the beat-frequency-resolved 2D-ES

In conventional 2D spectra, it is difficult to find the beating signal due to quantum coherence against the backdrop of the strong signal of the stationary component. To extract the beating signal from amidst the stationary component, we analyze the beat-frequency-resolved 2D spectra from a series of time-resolved 2D spectra [19-21].

The method to obtain conventional 2D spectra has been previously reported [11, 22]. First, we measure the spectra of the four-wave mixing (FWM) signal while scanning the coherent time $\tau$ and waiting time $T$:

$$S(\tau, \omega_D, T) = |E_{sig}(\tau, \omega_D, T)|^2 + E_{sig}(\tau, \omega_D, T)E_{LO}(\omega_D)^* \exp[i\omega_D(t_{LO} + \tau)] + c.c. \quad (1)$$

The first and second terms are the homodyne and heterodyne signals, respectively. The contribution of the local oscillator is subtracted by the chopper system described in Sec. 2-2. $E_{sig}$ and $E_{LO}$ are the electric fields of the FWM signal and the LO light, respectively; $\omega_D$ is the frequency in the spectra of the signal light; and $t_{LO}$ is the time delay of the LO pulse from the probe light $E_3$ as shown in Fig. 1(e) in the main text.

The measured spectra are inverse-Fourier transformed at each waiting time:

$$\tilde{S}(\tau, t, T) = \tilde{S}_{sig}(\tau, t, T) + \tilde{A}(t - t_{LO} - \tau, T) + \tilde{A}^*(t + t_{LO} + \tau, T). \quad (2)$$

The first term, which is the contribution of the homodyne signal, appears close to $t = 0$, and is blocked in the following Fourier transformation to remove the homodyne signal. Then we obtain:

$$E_+(\tau, \omega_D, T) = A(\tau, \omega_D, T) \exp[i\omega_D(t_{LO} + \tau)] \quad (3\text{-a})$$

and:

$$E_-(\tau, \omega_D, T) = A^*(\tau, \omega_D, T) \exp[-i\omega_D(t_{LO} + \tau)]. \quad (3\text{-b})$$

These are the Fourier-transformed spectra of the second and third terms in Eq. (2), respectively. Finally, we obtain the spectra of the heterodyne signal normalized by the LO spectrum $S_{LO}(\omega_D)$ as follows:

$$|E_{sig}(\tau, \omega_D, T)|^2 = E_+(\tau, \omega_D, T)E_-(\tau, \omega_D, T)/S_{LO}(\omega_D). \quad (4)$$

Then, the rephasing and non-rephasing 2D spectra $\tilde{I}(\omega_E, \omega_D, T)$ are calculated by taking the Fourier transform of Eq. (4) at each frequency $\omega_D$ in the region of $\tau > 0$ and $< 0$, respectively, where $\omega_E$ is the angular excitation frequency.

The 2D spectra are divided into 4 THz squares to obtain the pump-probe time profiles for each excitation and detection frequency. An example of the time profiles is shown in Fig.

3(a) in the main text. Then, the time profiles are Fourier transformed with respect to the waiting time $T$ to obtain $I(\omega_E, \omega_D, \omega_{beat})$, where $\omega_{beat}$ is the angular beat frequency. The three-dimensional spectrum $I(\omega_E, \omega_D, \omega_{beat})$ is shown in Supplemental Fig. S10. The horizontal axis is the beat frequency $E_{beat}$, and the vertical axis is the detection frequency $E_{det}$. In Supplemental Fig. S10, we show the spectra at the eight excitation frequencies, $E_{exc} = 496, 504, 512, 520, 528, 536, 544$, and 552 THz. We can easily see that the intensity of a specific beat frequency significantly depends on both the excitation and detection frequencies. For example, component 1 at the beat frequency of 16.5 THz appears at the low excitation frequency but, on the other hand, component 5 at $E_{beat} = 46.0$ THz appears at a high $E_{exc}$. Note that two distinguishable peaks (components 3 and 4) with an interval of 1.2 THz (40 cm$^{-1}$) show different dependencies on both the excitation and detection frequencies.

After we extract the intensity at a specific beat frequency and rearrange it as a 2D map with respect to $E_{exc}$ and $E_{det}$, the beat-frequency-resolved 2D spectra can be obtained, as shown in Fig. 4 in the main text.

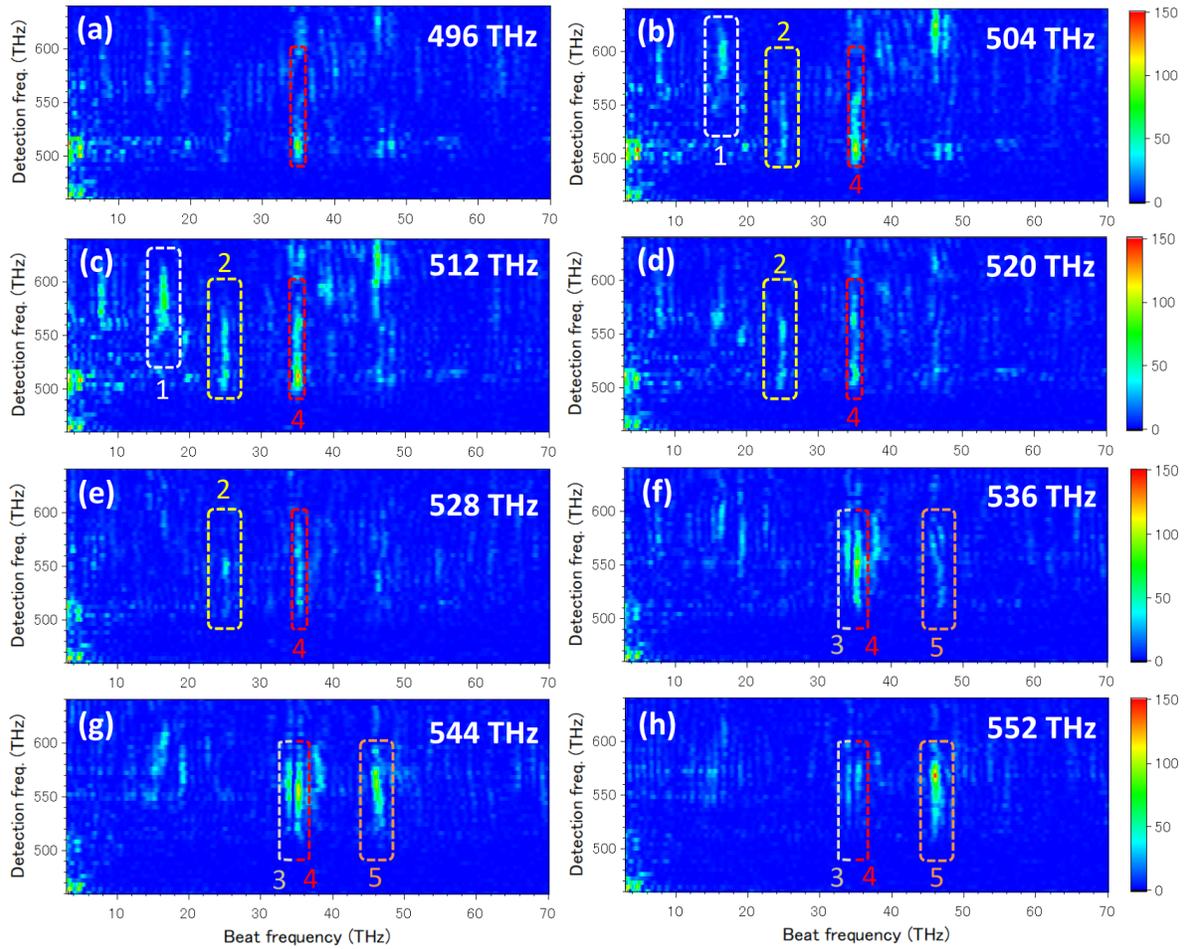

Supplemental Fig. S10. Beat-frequency spectra for the rephasing signal of RasM with respect to the excitation and detection frequencies. The horizontal axis shows the beat frequency. The vertical axis shows the detection frequency in the raw 2D spectra. Figures (a), (b), (c), (d), (e), (f), (g), and (h) show the spectra at the excitation frequencies of 496, 504, 512, 520, 528, 536, 544, and 552 THz, respectively. The dashed rectangles 1, 2, 3, 4, and 5 mark the components with beat frequencies of $E_{beat}$ = 16.5, 25.0, 34.0, 35.2, and 46.0 THz, respectively.

## 10. Rephasing pathways and related Feynman diagrams

Supplemental Fig. 11, 12, and 13 summarize the reaction pathways and related Feynman diagrams considered in this study.

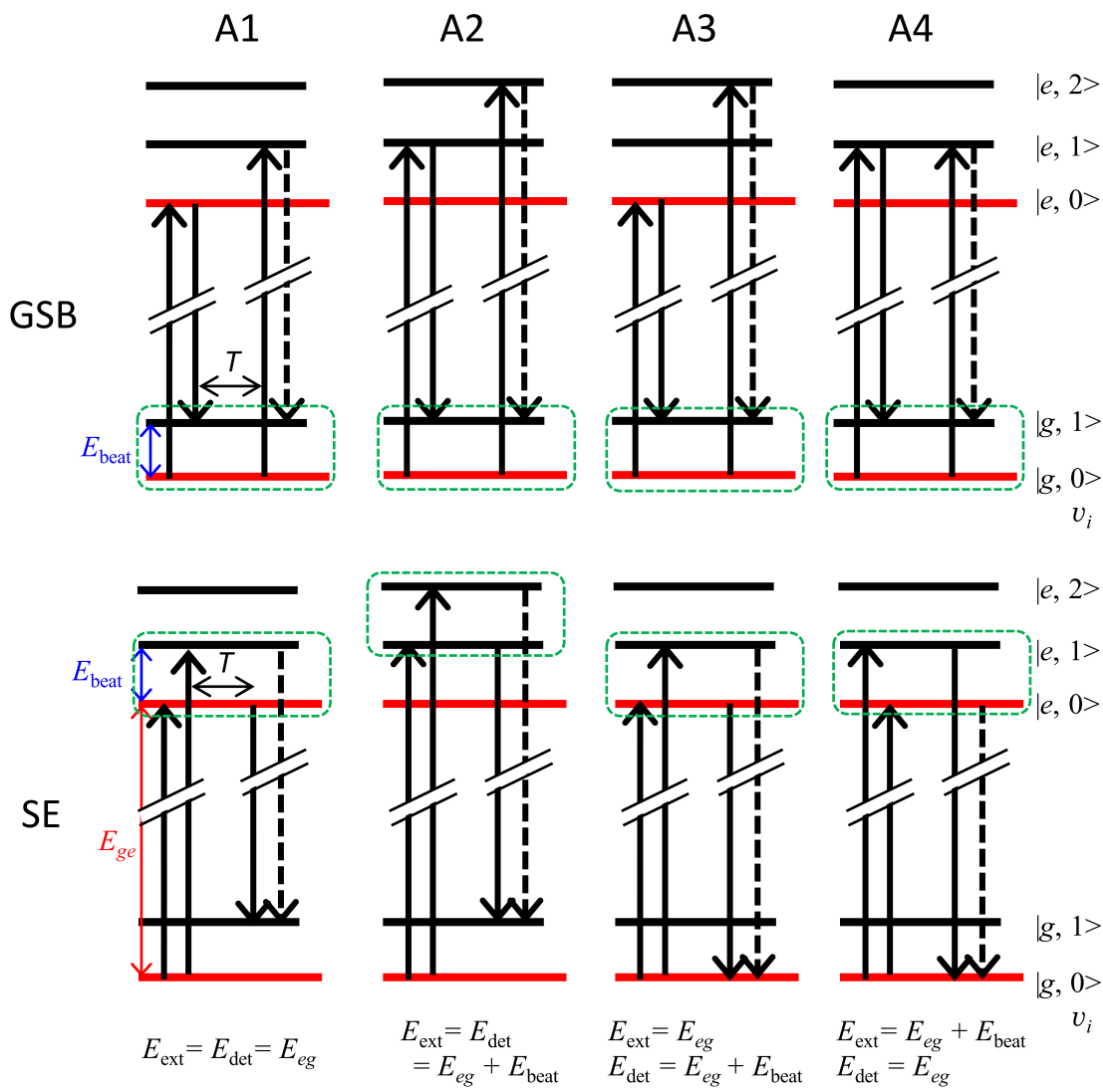

Supplemental Fig. S11. Rephasing pathways of the GSB and SE processes for the peaks A1, A2, A3, and A4. The beat frequency $E_{beat}$ is the energy spacing between the coherently excited vibronic states indicated by the green dashed rectangles. $E_{ge}$ is the energy of the first excited electronic state $|e\rangle$. The red horizontal lines show the electronic levels. The upper and lower rows show the GSB and SE pathways, respectively.

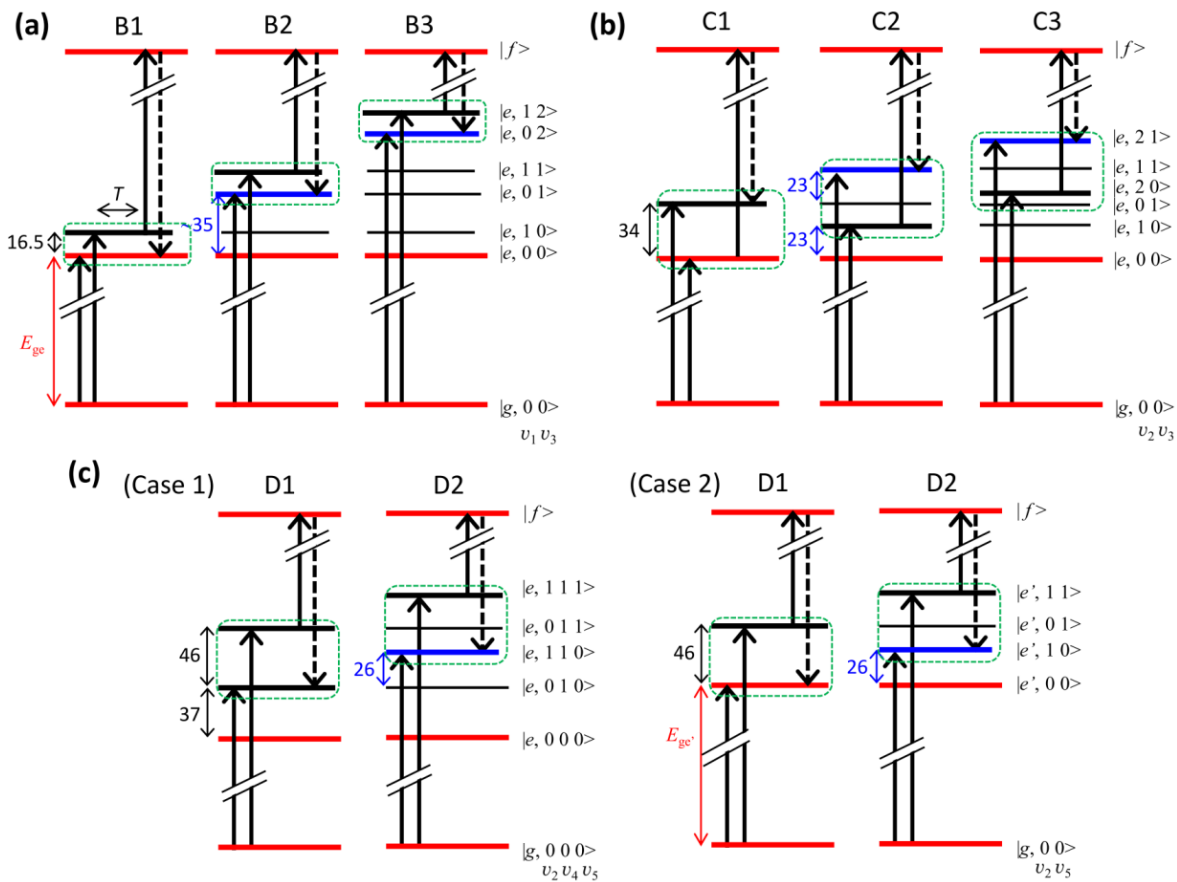

Supplemental Fig. S12. (a), (b), and (c) Rephasing pathways of the ESA processes for the peaks B, C, and D, respectively. The coherently excited vibronic states indicated by the green dashed rectangles result in the vibrational coherence. $E_{ge}$ is the energy of the first excited electronic state $|e\rangle$. The numbers show the vibrational energies in terahertz. The red horizontal lines show the electronic levels. The blue horizontal lines show the vibrationally excited levels contributing to the vibrational progression along the excitation frequency axis in the 2D maps. Cases 1 and 2 for peak D show the pathways with the first $|e\rangle$ and the second $|e'\rangle$ excited electronic states, respectively.

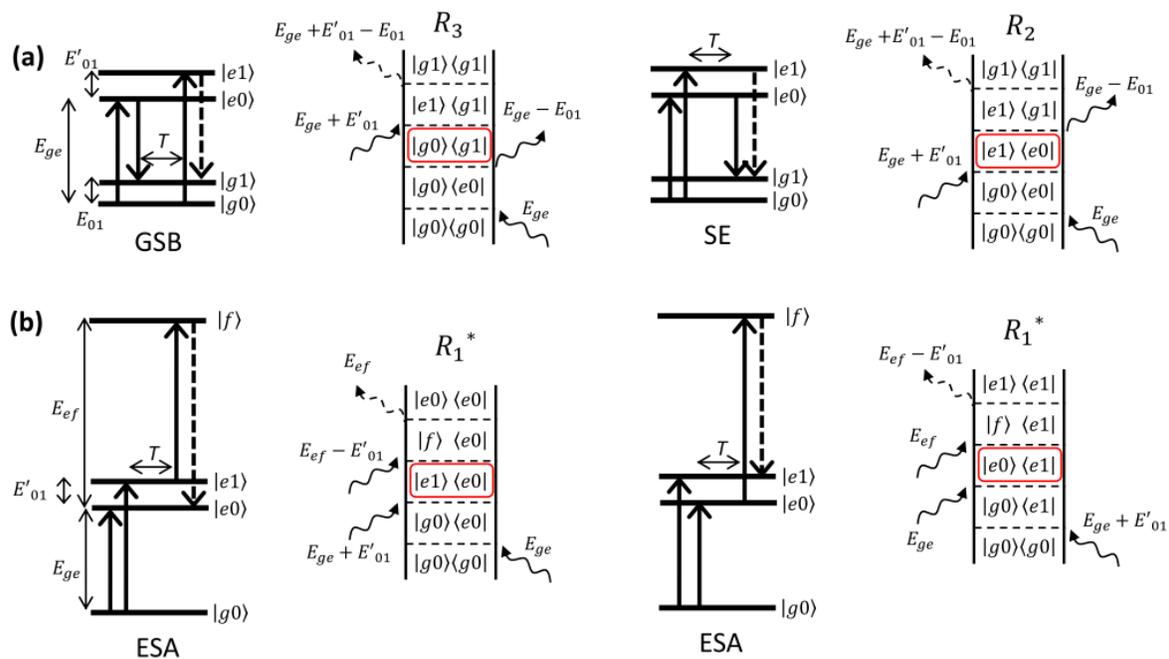

Supplemental Fig. S13. Double-sided Feynman diagrams representing the rephasing pathways of the four-wave mixing processes. (a) Vibronic state model for diagonal peak. (b) Vibronic state model for off-diagonal peak ($E_{\text{exc}} < E_{\text{det}}$). The red rectangles indicate the vibrational coherence observed for waiting time $T$.

Table S2. List of the FWM signals of the beat-frequency-resolved 2D spectra.

The center frequency and widths of excitation and detection axes are shown. The letters A to D in the first column correspond to those in Fig. 4.

|    | $E_{beat}$ (THz) | $E_{exc}$ (THz) Center | Width | $E_{det}$ (THz) Center | Width |
|----|------|--------|-------|--------|-------|
| B1 | 16.5 | 510 | 12 | 580 | 55 |
| B2 |      | 542 | 15 | 590 | 40 |
| B3 |      | 581 | 14 | 595 | 20 |
| A1 | 25.0 | 512 | 27 | 511 | 16 |
| A3 |      | 518 | 26 | 552 | 24 |
| C1 | 34.0 | 542 | 14 | 565 | 40 |
| C2 |      | 567 | 10 | 560 | 35 |
| C3 |      | 588 | 7.5 | 570 | 20 |
| A1 | 35.2 | 508 | 17 | 511 | 23 |
| A2 |      | 540 | 13 | 555 | 35 |
| A3 |      | 510 | 16 | 560 | 17 |
| D1 | 46.0 | 547 | 12 | 568 | 40 |
| D2 |      | 573 | 17 | 572 | 40 |

Table S3. Electronic and vibrational energies observed in the beat-frequency-resolved 2D spectra.

The "Beat" and "Excitation" rows show the vibrational modes contributing to the beating signal and the vibrational progression along the excitation frequency axis, respectively. The letters A to D indicate the component of the beat-frequency-resolved 2D maps shown in Fig. 4.
*In Cases 1 and 2 shown in Fig. 9(c), vibrational mode $\nu_4$ and electronic state $e'$ are adopted, respectively.

| Electronic state | $e$ | $e'^{*}$ | | | |
|---|---|---|---|---|---|
| Energy [THz (cm$^{-1}$)] | 512 (17,070) | 547 (18,250) | | | |
| Vibrational mode | $\nu_1$ | $\nu_2$ | $\nu_3$ | $\nu_4^{*}$ | $\nu_5$ |
| Energy [THz (cm$^{-1}$)] | 17 (550) | 23–26 (770–870) | 35 (1170) | 37 (1230) | 46 (1530) |
| Beat | B | A | A, C | | D |
| Excitation | | C, D | B | D | |

## 11. Data analysis for the non-rephasing signals

We also perform an analysis of the non-rephasing signal. Supplemental Fig. S14 shows the three-dimensional spectrum $S(\omega_E, \omega_D, \omega_{beat})$. Unlike the rephasing signal, the characteristic component is not found, except for component 4 at $\omega_{beat} = 46.0$ THz. Supplemental Fig. S15 shows the beat-frequency-resolved 2D spectra. Only a small number of significant components are found in the spectra. We are still considering why the non-rephasing signal does not have rich information on the vibronic coherence.

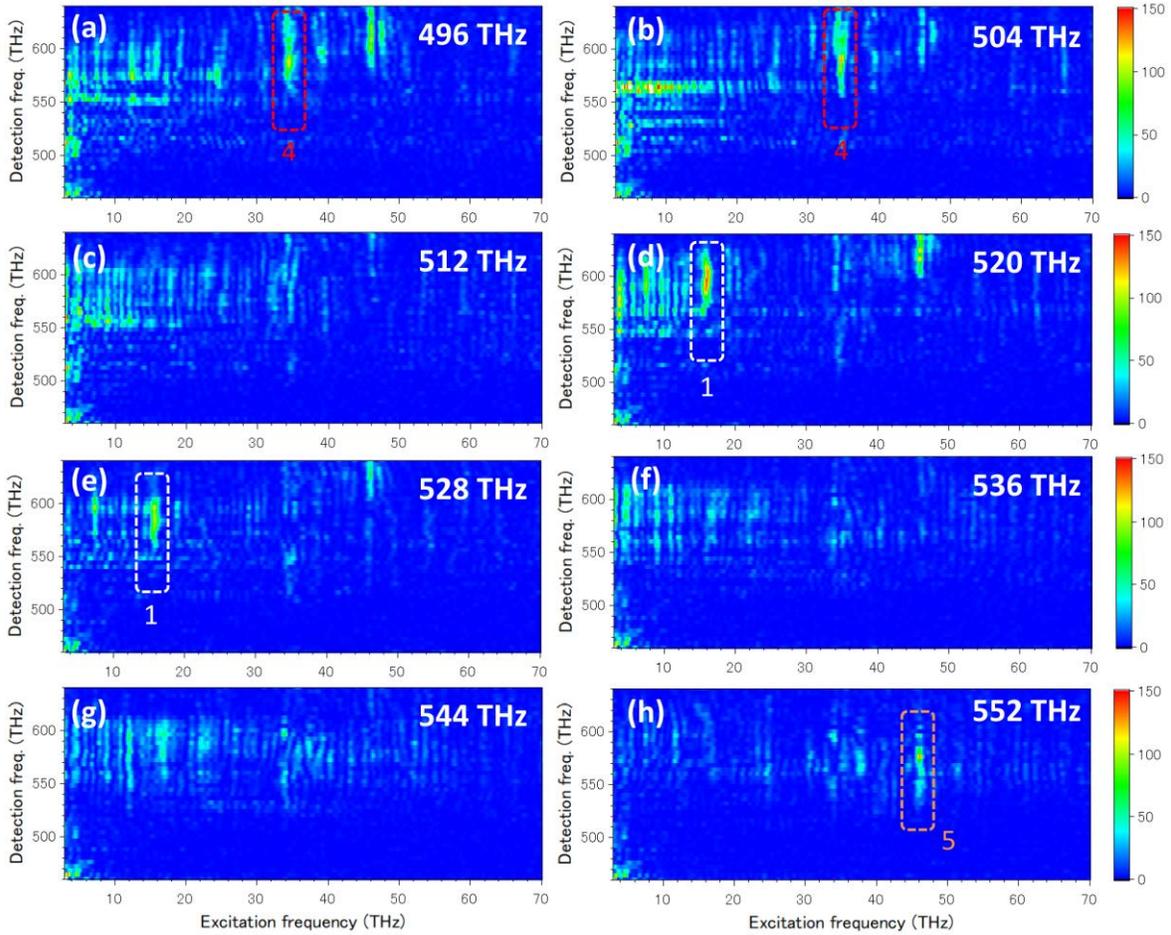

Supplemental Fig. S14. Beat-frequency spectra for the non-rephasing signal of RasM with respect to the excitation and detection frequencies. The horizontal axis shows the beat frequency. The vertical axis shows the detection frequency in the raw 2D spectra. Figures (a), (b), (c), (d), (e), (f), (g), and (h) show the spectra at the excitation frequencies of 496, 504, 512, 520, 528, 536, 544, and 552 THz, respectively. The dashed rectangles 1, 4, and 5 mark the components with beat frequencies of $E_{beat}$ = 16.5, 35.0, and 46.0 THz, respectively.

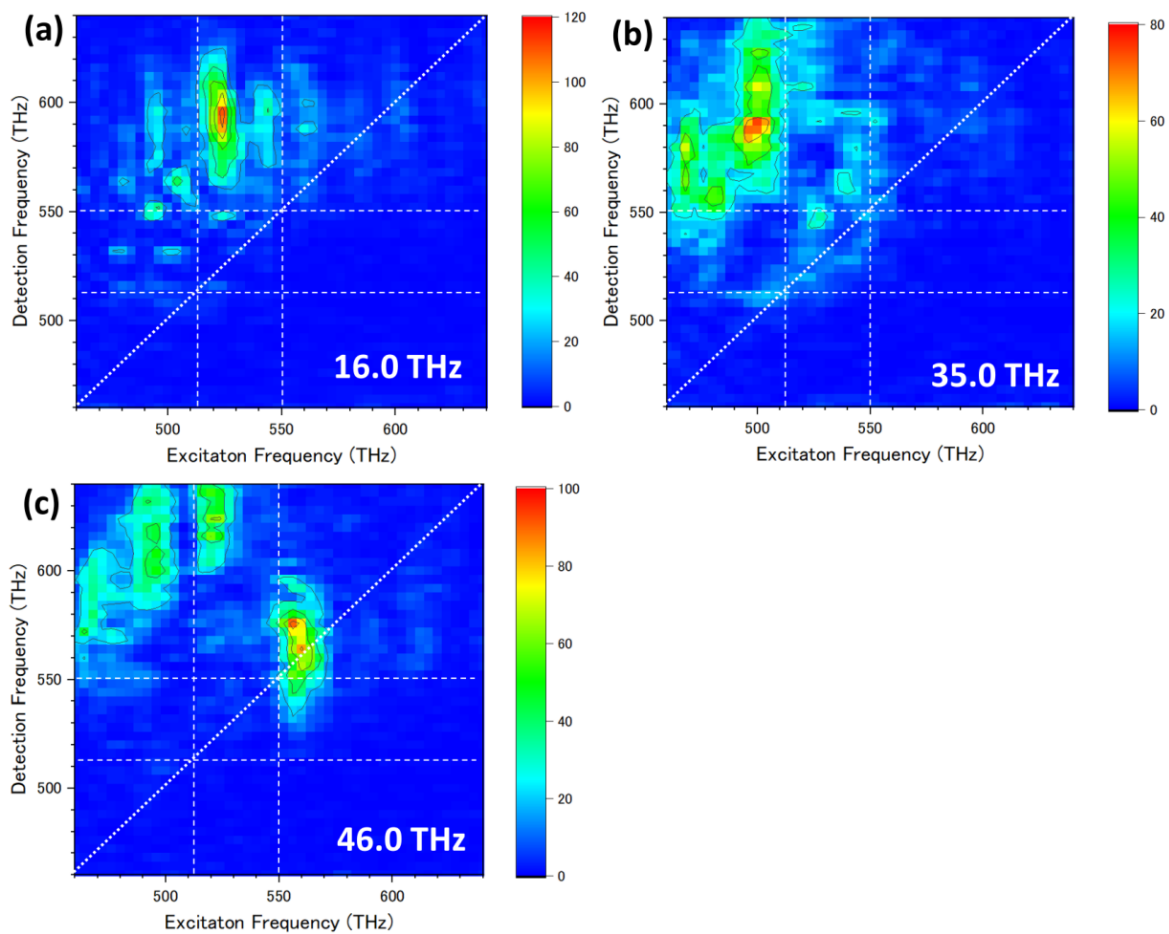

Supplemental Fig. S15. Beat-frequency-resolved 2D spectra for the non-rephasing signal of RasM. The 2D spectra (a), (b), and (c) are extracted from the total 2D spectra at beat frequencies of 16.0, 35.0, and 46.0 ± 1 THz, respectively.